\documentclass[aps,superscriptaddress,amsfonts,amssymb,amsmath,twocolumn,floatfix]{revtex4}

\usepackage[USenglish]{babel}
\usepackage{epsfig}
\usepackage{graphics}
\usepackage{graphicx}
\usepackage{amsmath}
\usepackage{color}
\usepackage{comment}
\usepackage{slashed}
\usepackage{siunitx}
\usepackage{hyperref}

\usepackage{soul}
\usepackage[normalem]{ulem}
\usepackage{xcolor}
\usepackage[makeroom]{cancel}

\hypersetup{
    colorlinks=true,
    linkcolor=blue,
    citecolor=blue,
    filecolor=magenta,
    urlcolor=cyan,
    pdfpagemode=FullScreen,
    }

\date{\today}

\begin{document}

\title{A simple gravitational self-decoherence model}

\author{Gabriel H. S. Aguiar}
\email{ghs.aguiar@unesp.br}
\affiliation{Universidade Estadual Paulista, Instituto de F{\' i}sica Te{\' o}rica, Rua Dr. Bento Teobaldo Ferraz 271, Bl. II, 01140-070, S{\~ a}o Paulo, S{\~ a}o Paulo, Brazil}

\author{George E. A. Matsas}
\email{george.matsas@unesp.br}
\affiliation{Universidade Estadual Paulista, Instituto de F{\' i}sica Te{\' o}rica, Rua Dr. Bento Teobaldo Ferraz 271, Bl. II, 01140-070, S{\~ a}o Paulo, S{\~ a}o Paulo, Brazil}


\begin{abstract}
    One of the most significant debates of our time is whether our macroscopic world (i)~naturally emerges from quantum mechanics or (ii)~requires new physics. We argue for the latter and propose a simple gravitational self-decoherence mechanism. For this purpose, we postulate the existence of a Heisenberg cut such that particles with masses~$m$ much smaller and larger than a critical mass~$M_\text{C}$ (of the order of the Planck mass~$M_\text{P}$) would be necessarily treated according to quantum and classical rules, respectively. Our effective model is designed to capture the new physics that free quantum particles would experience as their masses approach~$M_\text{C}$. The purity loss for free quantum particles is evaluated and shown to be highly inefficient for quantum particles with~$m \ll M_\text{C}$ but very effective for those with~$m \sim M_\text{C}$. The physical picture behind it is that coherence would (easily) leak from heavy enough particles to (non-observable) spacetime quantum degrees of freedom. Finally, we contextualize our proposal with state-of-the-art experiments and show how it can be tested in a future Stern-Gerlach-like experiment.
\end{abstract}

\maketitle

\section{Introduction}
\label{sec_introduction}

One of the most intriguing puzzles of modern physics is harmonizing the world we experience with quantum mechanics~(QM). The prevailing view is that the macroscopic world would probably emerge from the elementary blocks of nature, which~QM superbly describes. From this perspective, there would be no conceptual barrier, e.g., to putting bodies with large masses in a spatial quantum superposition, although technical difficulties would make it unfeasible in practice. This viewpoint splits into those that do not assign gravity any significant role (see, e.g., Ref.~\cite{2022-zurek} and references therein) and those that concede some protagonism to gravity. Some models in the latter group are entirely rooted in general relativity~\cite{2013-blencowe, 2013-anastopoulos, 2021-asprea}, while others go beyond general relativity and realize quantum-gravity degrees of freedom expected to appear at the Planck scale. Still, both parties in the latter group share the common ground that~QM would apply to particles with macroscopic masses. For instance, assuming a thermal bath of primordial gravitons (massless spin-2 particles) at the temperature~$\mathcal{T}_\text{g} \sim 1~\si{\kelvin}$, Blencowe obtains that a system of {\em some grams}, with {\em an Avogadro number of atoms} (with all of them superposed in the ground and some excited state, differing of~$1~\si{\electronvolt}$), would decohere in a time scale of~$10^{- 2}~\si{\second}$~\cite{2013-blencowe}. In turn, inspired by Hawking~\cite{1982-hawking}, Ellis, Mohanty, and Nanopoulos modeled spacetime quantum degrees of freedom as a gas of wormholes~\cite{1989-ellis}, concluding that systems with {\em an Avogadro number of particles} could keep coherence up to~$10^{- 7}~\si{\second}$. (More recently, Arzano et al. formulated a model of fundamental decoherence assuming a quantum spacetime where time and space coordinates would not commute at the Planck scale~\cite{2023-arzano}.)

\begin{figure}[b]
    \includegraphics[width = 40mm]{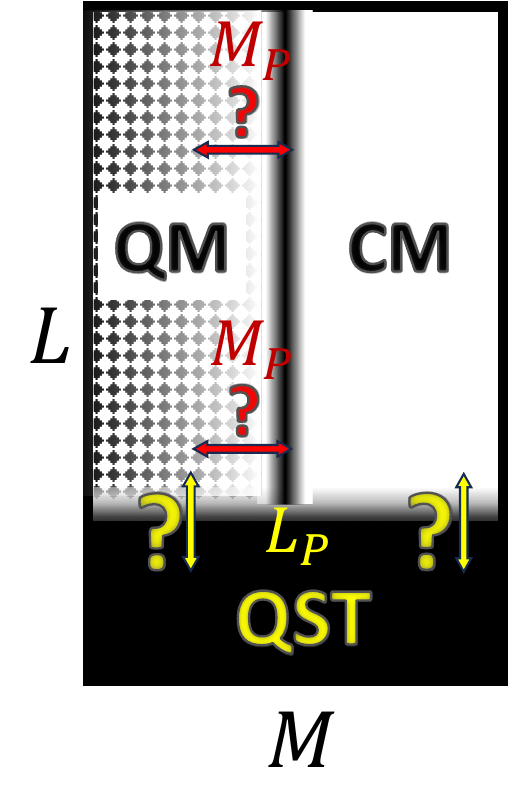}
    \caption{The black strip at~$L < L_\text{P} \sim 10^{- 35}~\si{\meter}$ veils new physics necessary to probe {\em proper} distances smaller than the Planck length~$L_\text{P}$, where the spacetime would be described by some (still unknown) quantum spacetime theory~(QST). This would jeopardize the existence (or at least, description) of classical particles with lengths~$l \ll L_\text{P}$ and quantum particles with wavelengths~$\lambdabar \ll L_\text{P}$. In our model, particles with masses much smaller, $m \ll M_\text{P}$, and larger, $m \gg M_\text{P}$, than the Planck mass, $M_\text{P} \sim 10^{- 5}~\si{\gram}$, would necessarily be governed by~QM and~CM, respectively. The vertical black strip indicates the corresponding Heisenberg cut. Particles with masses~$m \sim M_\text{P}$ would require new physics to be described. According to our model, quantum particles, $m \lesssim M_\text{P}$, would experience a significant gravitational self-decoherence as their masses approach the Heisenberg cut, $m \to M_\text{P}$.}
    \label{fig_manifesto}
\end{figure}

On the other hand, a less frequent yet equally significant standpoint is that usual~QM should be altered to account for the macroscopic world. Similarly to the prevailing case, this vision also splits into those who do not assign any unique role to gravity (as the Ghirardi-Rimini-Weber spontaneous collapse model~\cite{2003-bassi}) and those who do (see, e.g., Refs.~\cite{2013-bassi, 2017-bassi, 2002-anastopoulos} for comprehensive reviews). In the latter group, the Di{\' o}si-Penrose~(DP) gravitational self-decoherence model stands out for being extensively studied~\cite{1996-penrose, 1998-penrose, 1989-diosi}. It was initially conjectured that its free parameter would be~$R_0 \approx 10^{- 15}~\si{\meter}$, but recent experiments have pushed it to~$R_0 > 10^{- 10}~\si{\meter}$~\cite{2021-donadi}. It belongs to the set of models where quantum states experience phases of unitary Schr{\" o}dinger evolution followed by objective collapse periods. While the unitary evolution lasts eons for microscopic quantum systems, it would often be interrupted by objective collapse periods for macroscopic bodies. Although reckoning on new physics (by averaging metric fluctuations), the~DP model still assumes that macroscopic bodies would evolve quantum mechanically between the collapse events.

Here, we put forward a distinct viewpoint, where classical mechanics~(CM) and~QM would emerge on equal terms from some still unknown quantum spacetime theory~(QST) --- see Fig.~\ref{fig_manifesto}. New physics would appear at the Heisenberg cut for particles with masses of the Planck mass order, $M_\text{P} \sim 10^{- 5}~\si{\gram}$. In Sec.~\ref{sec_heisenberg}, we elaborate on this viewpoint. In Sec.~\ref{sec_review} we briefly review standard~QM applied to composed systems to be used in the subsequent section. In Sec.~\ref{sec_model}, we introduce our gravitational self-decoherence model. In Sec.~\ref{sec_free}, we apply the model to calculate the gravitational self-decoherence for free particles. In Sec.~\ref{sec_experiments}, we put the model in context concerning state-of-the-art experiments. In Sec.~\ref{sec_sg}, we show how the model can be tested in a Stern-Gerlach-like experiment. Our final comments appear in Sec.~\ref{sec_conclusions}.

\section{A case for a Heisenberg cut}
\label{sec_heisenberg}

Standard physical theories are written over classical spacetimes. The nonrelativistic Galilei spacetime demands clocks and rulers to be defined. In contrast, relativistic spacetimes only require clocks to be completely specified. Not only this but observables defined in relativistic spacetimes can all be evaluated using clocks only. We refer the reader to Ref.~\cite{2024-matsas} for details and the next two paragraphs for a summary.

To measure the length~$l$ of a rod in Minkowski spacetime, one can use the following protocol using three clocks. An inertial clock~$\mathcal{C}_1$ flies from the left to the right end of the rod, recording its trip-time interval~$\tau_1$. Immediately after, an inertial clock~$\mathcal{C}_2$ is sent back, recording its trip-time interval~$\tau_2$. Meanwhile, a clock~$\mathcal{C}_3$ records the time interval~$\tau_3$ that elapses from the departure of~$\mathcal{C}_1$ to the return of~$\mathcal{C}_2$. The time intervals~$\tau_1$, $\tau_2$, and~$\tau_3$ properly combine such that the rod's length can be expressed as
\begin{equation}
    l
    =
    \frac{\sqrt{(\tau_1^2 + \tau_2^2 - \tau_3^2)^2 - 4 \tau_1^2 \tau_2^2}}{2 \tau_3}.
    \label{UE}
\end{equation}
(Let us emphasize that this result will hold irrespective of how quickly the inertial clocks~$\mathcal{C}_1$ and~$\mathcal{C}_2$ move, for any rods in Minkowski spacetime and any small enough rods in general relativistic spacetimes.) Hereafter, $c = 1$, and all distances are measured in seconds (or, equivalently, light seconds). We see, hence, that distances can be measured only with clocks {\em in relativistic spacetimes}. (See Sec.~{\em ``Time to rule us all in relativistic spacetimes"} of Ref.~\cite{2024-matsas} for more details.) Note that the protocol above is invalid in Galilei spacetime, reflecting the need for independent standards of space and time to express observables in nonrelativistic spacetimes.

Furthermore, mass, i.e., a body's ``attractive power," can also be measured only with clocks in relativistic spacetimes (assumed the equivalence principle) --- see Ref.~\cite{2024-matsas}. First, we recall that Isaac Newton expressed his gravitation law with~$G = 1$. He did not (nobody does) need~$G$ for any prediction. It was not until 1873, i.e., 150 years after Newton's death, that~$G$ was introduced. The very year~$G$ was introduced, James Maxwell wrote in the Preliminaries of his {\it A Treatise on Electricity and Magnetism}: {\em ``If, as in the astronomical system, the unit of mass is defined with respect to its attractive power, the dimensions of~$[\text{mass}]$ are~$[\text{length}]^3 / [\text{time}]^2$."} Indeed, $G$ is a conversion factor of units, from~$\si{\meter}^3 / \si{\second}^2$ to~$\si{\kilo\gram}$, with the kilogram being introduced during the French revolution for commercial reasons. (After the insertion of~$G$, the ``attractive power" of a body is given by~$G M$.) Combining this discussion with the previous paragraph, we have that mass can be measured only with clocks {\em in relativistic spacetimes}, and has dimension of~$\si{\second}$ (in units where~$c = G = 1$). (See Sec.~{\em ``Two units to rule us all in Galilei spacetime"} of Ref.~\cite{2024-matsas} for more details).

Discussions of fundamental issues should knock out superfluous structures. Thus, we assume that relativistic spacetimes are endowed only with clocks, which are just enough to evaluate all observables. Hereafter, $c = G = 1$, and all observables are measured in seconds.

Spacetime specification requires clocks to be (i)~pointlike and possess (ii)~well-defined worldlines. Condition~(i) is a mathematical idealization since~QM poses a physical barrier to building arbitrarily small clocks. A more realistic account would describe bona fide clocks as having tiny quantum clockworks to approximate~(i) and being encapsulated in classical containers to satisfy~(ii). Unfortunately, such (realistic) clocks will not be able to measure time intervals (and, thus, space distances) at the Planck scale, impacting~QM's~domain of applicability. To see it, let us consider the most straightforward instance in~QM.

Take a free particle with mass~$m$, and corresponding reduced Compton wavelength~$\lambdabar$, described by Dirac or Klein-Gordon equations:
\begin{equation}
    (i \slashed{\partial} - \lambdabar^{- 1}) \psi
    =
    0, 
    \quad
    (\Box + \lambdabar^{- 2}) \phi
    =
    0.
\end{equation}
To resolve a distance~$\lambdabar$, clocks should resolve time intervals
\begin{equation}
    \Delta t
    <
    \lambdabar
    \label{A1}
\end{equation}
[to measure the time interval taken by a light ray to cover the~$\lambdabar$ distance or use Eq.~\eqref{UE}]. To resolve such time intervals, the clockwork must have a mass
\begin{equation}
    M_\text{CW}
    \geq
    \hbar / (2 \Delta t)
    \label{A2}
\end{equation}
(see, e.g., \S~5.3 of Ref.~\cite{2010-schumacher}). The clockwork is assumed to be encapsulated in a classical container with radius~$l_\text{clock}$ to comply with condition~(ii) above. The Schwarzschild radius associated with the total clock mass, $R_\text{Sc} = 2 M_\text{clock}$, will, then, satisfy the constraint
\begin{equation}
    R_\text{Sc}
    >
    \hbar / \Delta t
    >
    L_\text{P}^2 / \lambdabar,
\end{equation}
where we have used Eqs.~\eqref{A1} and~\eqref{A2}. This must be compared with the clock size which must satisfy
\begin{equation}
    l_\text{clock}
    <
    \lambdabar.
\end{equation}
Hence, to measure lengths~$\lambdabar \lesssim L_\text{P}$, clocks should be so small as to collapse into black holes: $l_\text{clock} < R_\text{Sc}$.

Also, using gravity first to determine~$m$ and then inferring~$\lambdabar$ is not an option either. To measure mass with gravity, we must both (a)~set the~CoM~of~$m$ at some known distance~$l$ from a test particle and (b)~make sure that~$m$ lies at rest. However, being~$m$ a quantum particle, position and momentum cannot be simultaneously measured since they are incompatible variables. Once the particle's position (momentum) is measured, we lose total knowledge of the momentum (position), ruining the gravitational measurement. Eventually, this problem can be traced back to our ignorance about how quantum particles curve spacetime.

We are confronted, thus, with the extraordinary situation where not only (I)~present spacetime descriptions should fail as one probes distances~$l \ll L_\text{P}$, but also (II)~usual~QM would not apply to particles with masses~$m \gg M_\text{P}$ (reduced Compton wavelengths~$\lambdabar \ll L_\text{P}$). By this token, we consider an alternative picture imposing a Heisenberg cut at a mass~$M_\text{C} \sim M_\text{P}$ such that particles with masses~$m \ll M_\text{C}$ and~$m \gg M_\text{C}$ should be necessarily treated using~QM and~CM, respectively. The Heisenberg cut would segregate the quantum and classical realms --- see Fig.~\ref{fig_manifesto}. In contrast, {\em particles with masses close to the Heisenberg cut, $m \sim M_\text{C}$, should obey some new physics}.

For such new physics, we propose an elementary gravitational self-decoherence mechanism designed for particles with masses~$m < M_\text{C}$ that protects the quantum coherence of particles with~$m \ll M_\text{P}$ (in agreement with usual~QM) but efficiently decoheres particles with masses~$m \to M_\text{C} \sim M_\text{P}$. We emphasize that, from our perspective, particles with~$m \gg M_\text{P}$ should not admit {\em any} quantum-inspired treatment.

\section{Brief review on composed quantum systems}
\label{sec_review}

Since our model demands massive arrangements to be tested, let us start with a nonrelativistic quantum system composed of~$N$ scalar particles with masses~$m_j$ ($j = 1, \ldots, N$). This is described by a wavefunction~$\psi_N(\{\mathbf{r}_j\}, t) \in \bigotimes_{\ell = 1}^N L^2(\mathbb{R}^3)$, normalized by
\begin{equation}
    \left(\prod\limits_{\ell = 1}^N \int \! d^3 \mathbf{r}_\ell\right) \psi_N^*(\{\mathbf{r}_j\}, t) \psi_N(\{\mathbf{r}_j\}, t)
    =
    1,
    \label{N1}
\end{equation}
and satisfying the Schr{\" o}dinger equation
\begin{equation}
    i \hbar \; \partial_t \psi_N(\{\mathbf{r}_j\}, t)
    =
    H_N(\{\mathbf{r}_j\}, t) \psi_N(\{\mathbf{r}_j\}, t).
    \label{SE1}
\end{equation}
The Hamiltonian~$H_N(\{\mathbf{r}_j\}, t)$ consists of the kinetic term
\begin{equation}
    K_N(\{\mathbf{r}_j\})
    \equiv
    - \sum\limits_{\ell = 1}^N \frac{\hbar^2}{2 m_\ell} \nabla_{\mathbf{r}_\ell}^2,
    \label{K1}
\end{equation}
some external potential~$V_N(\{\mathbf{r}_j\}, t)$, and an internal potential~$U_N(\{|\mathbf{r}_j - \mathbf{r}_{j'}|\})$ with~$j, j' = 1, \ldots, N$ ($j' \neq j$).

Now, let us define the coordinates of the~CoM and the relative ones as
\begin{equation}
    \mathbf{r}
    \equiv
    \sum\limits_{\ell = 1}^N \frac{m_\ell}{m} \mathbf{r}_\ell
    \quad
    \text{and}
    \quad
    \Tilde{\mathbf{r}}_k
    \equiv
    \mathbf{r}_k - \mathbf{r}
    \quad
    (k = 1, \ldots, N - 1),
    \label{C1}
\end{equation}
respectively, where~$m \equiv \sum_{\ell = 1}^N m_\ell$. In terms of them, the particle coordinates can be expressed as
\begin{equation}
    \mathbf{r}_k
    =
    \mathbf{r} + \Tilde{\mathbf{r}}_k
    \quad
    (k = 1, \ldots, N - 1),
    \quad
    \mathbf{r}_N
    =
    \mathbf{r} - \sum\limits_{\ell = 1}^{N - 1} \frac{m_\ell}{m_N} \Tilde{\mathbf{r}}_\ell.
    \label{C2}
\end{equation}
Now, using Eq.~\eqref{C1}, we cast Eq.~\eqref{K1} as (see, e.g., \cite{2014-giulini})
\begin{equation}
    K_N(\{\mathbf{r}_j\})
    =
    K(\mathbf{r}) + \Tilde{K}(\{\Tilde{\mathbf{r}}_k\}),
    \label{K2}
\end{equation}
where~$K(\mathbf{r}) \equiv - \hbar^2 / (2 m) \nabla_\mathbf{r}^2$ and
\begin{equation}
    \Tilde{K}(\{\Tilde{\mathbf{r}}_k\})
    \equiv
    - \frac{\hbar^2}{2} \sum\limits_{\ell = 1}^{N - 1} \sum\limits_{\ell' = 1}^{N - 1} \left(\frac{\delta_{\ell \ell'}}{m_\ell} - \frac{1}{m}\right) \mathbf{\nabla}_{\Tilde{\mathbf{r}}_\ell} \cdot \mathbf{\nabla}_{\Tilde{\mathbf{r}}_{\ell'}}.
\end{equation}
If the external potential is uniform over the particles, it can be cast as a simple function of the~CoM~coordinates:
\begin{equation}
    V_N(\{\mathbf{r}_j\}, t)
    \to
    V(\mathbf{r}, t).
    \label{V}
\end{equation}
Similarly, it follows from Eq.~\eqref{C2} that the internal potential only depends on the relative coordinates:
\begin{equation}
    U_N(\{|\mathbf{r}_j - \mathbf{r}_{j'}|\})
    \to
    \Tilde{U}(\{\Tilde{\mathbf{r}}_k\}).
    \label{U1}
\end{equation}
Using Eqs.~\eqref{K2} and~\eqref{V}-\eqref{U1}, we cast the Hamiltonian as
\begin{equation}
    H_N(\{\mathbf{r}_j\}, t)
    =
    H(\mathbf{r}, t) + \Tilde{H}(\{\Tilde{\mathbf{r}}_k\}),
    \label{H1}
\end{equation}
where
\begin{eqnarray}
    H(\mathbf{r}, t)
    &\equiv&
    K(\mathbf{r}) + V(\mathbf{r}, t),
    \label{H2}
    \\
    \Tilde{H}(\{\Tilde{\mathbf{r}}_k\})
    &\equiv&
    \Tilde{K}(\{\Tilde{\mathbf{r}}_k\}) + \Tilde{U}(\{\Tilde{\mathbf{r}}_k\}).
\end{eqnarray}
Since Eqs.~\eqref{K2} and~\eqref{U1} always hold, the validity of Eq.~\eqref{H1} depends on the uniformity of~$V_N(\{\mathbf{r}_j\}, t)$. Assuming it, let us look for a solution of Eq.~\eqref{SE1} as
\begin{equation}
    \psi_N(\{\mathbf{r}_j\}, t)
    =
    C \; \psi(\mathbf{r}, t) \Tilde{\psi}(\{\Tilde{\mathbf{r}}_k\}, t)
    \quad
    (C = \text{const}),
\end{equation}
where~$\psi(\mathbf{r}, t) \in L^2(\mathbb{R}^3)$ and~$\Tilde{\psi}(\{\Tilde{\mathbf{r}}_k\}, t) \in \bigotimes_{\ell = 1}^{N - 1} L^2(\mathbb{R}^3)$ are the~CoM and relative wavefunctions, normalized as
\begin{eqnarray}
    &{}&
    \int \! d^3 \mathbf{r} \; \psi^*(\mathbf{r}, t) \psi(\mathbf{r}, t)
    =
    1,
    \label{N2}
    \\
    &{}&
    \left(\prod\limits_{\ell = 1}^{N - 1} \int \! d^3 \mathbf{r}_\ell\right) \Tilde{\psi}^*(\{\Tilde{\mathbf{r}}_k\}, t) \Tilde{\psi}(\{\Tilde{\mathbf{r}}_k\}, t)
    =
    1,
    \label{N3}
\end{eqnarray}
and satisfying the corresponding Schr{\" o}dinger equations:
\begin{eqnarray}
    i \hbar \; \partial_t \psi(\mathbf{r}, t) 
    &=&
    H(\mathbf{r}, t) \psi(\mathbf{r}, t),
    \\
    i \hbar \; \partial_t \Tilde{\psi}(\{\Tilde{\mathbf{r}}_k\}, t)
    &=&
    \Tilde{H}(\{\Tilde{\mathbf{r}}_k\}) \Tilde{\psi}(\{\Tilde{\mathbf{r}}_k\}, t),
\end{eqnarray}
respectively. Using Eqs.~\eqref{N2} and~\eqref{N3} in Eq.~\eqref{N1}, the normalization constant is fixed as~$C = (m_N / m)^{3 / 2}$~\cite{2014-giulini}. In what follows, we will focus on the~CoM~degrees of freedom for a particle at a uniform external potential, neglecting the relative degrees of freedom.

\section{A gravitational self-decoherence mechanism}
\label{sec_model}

Our model does not rely on any specific~QST. Instead, we {\em effectively} model the loss of quantum coherence for particles with~$m < M_\text{C}$ solely relying on (A)~the Heisenberg cut~$L_\text{C} \equiv \hbar / M_\text{C} \sim L_\text{P}$ and (B)~the particle features, given by its mass~$m$ and spreading~$\sigma$ (that defines the particle's initial state). For this purpose, we assume that the quantum particle has a ``virtual clone" with which it interacts gravitationally only. By ``clone," we mean it inherits the particle's properties and quantum state. These features guarantee that the particle and its clone will respond equally to external potentials, tracking each other's behavior. {\em Still, the particle and its clone are only permitted to interact gravitationally.} By labeling the clone as ``virtual," we mean that the clone does not exist as a physical entity (neither has to do with virtual particles of quantum field theory). Yet this is an instrumental concept to emulate the particle's loss of coherence to quantum gravitational degrees of freedom (whatever they are).

Although the reasoning favoring the Heisenberg cut in Sec.~\ref{sec_heisenberg} assumed relativistic spacetimes, our calculations will be performed in a reference frame where the particle's~CoM has negligible kinetic energy. This authorizes us to proceed with nonrelativistic~QM. By this token, the particle and clone will be considered altogether and described by the combined nonrelativistic wavefunction~$\Psi(\mathbf{r}, \Bar{\mathbf{r}}, t) \in L^2(\mathbb{R}^3) \otimes L^2(\mathbb{R}^3)$. This is normalized as
\begin{equation}
    \int \! d^3 \mathbf{r} \int \! d^3 \Bar{\mathbf{r}} \; \Psi^*(\mathbf{r}, \Bar{\mathbf{r}}, t) \Psi(\mathbf{r}, \Bar{\mathbf{r}}, t)
    =
    1,
\end{equation}
and satisfies the Schr{\" o}dinger equation
\begin{equation}
    i \hbar \; \partial_t \Psi(\mathbf{r}, \Bar{\mathbf{r}}, t)
    =
    [H(\mathbf{r}, t) + H(\Bar{\mathbf{r}}, t) + U(|\mathbf{r} - \Bar{\mathbf{r}}|)] \Psi(\mathbf{r}, \Bar{\mathbf{r}}, t)
    \label{SE2}
\end{equation}
with initial condition~$\Psi(\mathbf{r}, \Bar{\mathbf{r}}, 0) = \psi(\mathbf{r}, 0) \psi(\Bar{\mathbf{r}}, 0)$ (recall the clone inherits the particle's quantum state). Here, $H(\mathbf{r}, t)$ is given by Eq.~\eqref{H2} (and similarly for~$H(\Bar{\mathbf{r}}, t)$ replacing~$\mathbf{r}$ by the clone's coordinates~$\Bar{\mathbf{r}}$), and~$U(|\mathbf{r} - \Bar{\mathbf{r}}|)$ is the gravitational interaction potential between the particle and clone. To ensure consistency with the initial discussion that physical equations written over classical spacetimes should be meaningless below some distance scale, we introduce the cutoff parameter~$L_\text{C}$ in the Newtonian gravitational potential ($G = 1$)
\begin{equation}
    U(|\mathbf{r} - \Bar{\mathbf{r}}|)
    \equiv
    - m^2 / |\mathbf{r} - \Bar{\mathbf{r}}|_{L_\text{C}},
    \label{U2}
\end{equation}
by defining
\begin{equation}
    |\mathbf{r} - \Bar{\mathbf{r}}|_{L_\text{C}} 
    \equiv 
    [|\mathbf{r} - \Bar{\mathbf{r}}|^2 + L_\text{C}^2]^{1 / 2}.
\end{equation}
The size of~$L_\text{C} \sim L_\text{P}$ is defined in the~CoM~frame where the nonrelativistic calculations are carried out (not challenging the Lorentz invariance of the underlying relativistic spacetime). Let us note that, if~$U(|\mathbf{r} - \Bar{\mathbf{r}}|)$ was absent, Eq.~\eqref{SE2} would separate into two independent Schr{\" o}dinger equations
\begin{eqnarray}
    i \hbar \; \partial_t \psi(\mathbf{r}, t)
    &=&
    H(\mathbf{r}, t) \psi(\mathbf{r}, t),
    \\
    i \hbar \; \partial_t \psi(\Bar{\mathbf{r}}, t)
    &=&
    H(\Bar{\mathbf{r}}, t) \psi(\Bar{\mathbf{r}}, t),
\end{eqnarray}
where the particle and clone would evolve independently from each other. Only the gravitational potential~\eqref{U2} connects particle and clone, being also responsible for their entanglement. We should not forget, however, that, because the clone is a mathematical artifact of our effective model, we must eventually trace over its degrees of freedom. This procedure models the physical transfer of information from the particle to the (quantum) spacetime degrees of freedom (inaccessible in the laboratory).

It is conceptually pertinent (yet technically inconsequential) to cast Eq.~\eqref{U2} as
\begin{equation}
    U(|\mathbf{r} - \Bar{\mathbf{r}}|)
    =
    - \hbar^2 / (\lambdabar^2 |\mathbf{r} - \Bar{\mathbf{r}}|_{L_\text{C}}),
    \label{U3}
\end{equation}
where we recall that~$c = G = 1$ following the discussion in Sec.~\ref{sec_heisenberg}. Thus, Eq.~\eqref{U3} solely depends on the elementary spin scale ($\hbar$), the cutoff ($L_\text{C}$), and the particle itself ($\lambdabar$).

We must highlight the differences between our model and the model proposed by Filippo et al.~\cite{2002-filippo, 2003-filippo}, who examine a nonrelativistic gravitational self-decoherence model and apply it to systems composed of many particles, doubling the corresponding degrees of freedom. In their model, the cloned degrees of freedom are assumed to interact with all particles. In particular, potential~(4) for the~CoM in Ref.~\cite{2003-filippo} depends on the structure of the particle, while our corresponding potential~\eqref{U2} does not. As a consequence, Filippo et al.'s model becomes quite sensitive to the number of degrees of freedom, and decoherence would be observed for systems as light as with about~$10^{11}$ proton masses, i.e., $10^{- 13}~\si{\gram}$. In contrast, our model requires a single clone, scales with the mass of the~CoM, is insensitive to the number of constituents, and no~CoM~decoherence is predicted for systems with masses~$m \ll M_\text{C} \sim 10^{- 5}~\si{\gram}$.

\section{Decoherence of a free quantum particle}
\label{sec_free}

Let us begin considering a {\em free} particle, $V(\mathbf{r}, t) = 0$, initially described by a Gaussian wavepacket
\begin{equation}
    \psi(\mathbf{r}, 0)
    \equiv
    \frac{1}{(2 \pi \sigma^2)^{3 / 4}} \exp{\left(- \frac{\mathbf{r}^2}{4 \sigma^2}\right)}
    \label{W1}
\end{equation}
with standard deviation~$\sigma$. Then, $\Psi(\mathbf{r}, \Bar{\mathbf{r}}, t)$ describing the particle-clone pair will satisfy
\begin{equation}
    i \hbar \, \partial_t \Psi(\mathbf{r}, \Bar{\mathbf{r}}, t)
    \! = \!
    \left[- \frac{\hbar \lambdabar}{2} (\nabla_\mathbf{r}^2 + \nabla_{\Bar{\mathbf{r}}}^2) - \frac{\hbar^2}{\lambdabar^2 |\mathbf{r} - \Bar{\mathbf{r}}|_{L_\text{C}}}\right] \! \Psi(\mathbf{r}, \Bar{\mathbf{r}}, t)
    \label{SE3}
\end{equation}
with initial condition~$\Psi(\mathbf{r}, \Bar{\mathbf{r}}, 0) = \psi(\mathbf{r}, 0) \psi(\Bar{\mathbf{r}}, 0)$, and we recall that~$\hbar = L_\text{P}^2$ ($c = G = 1$).

Since we must ensure that the introduction of~$U(|\mathbf{r} - \Bar{\mathbf{r}}|)$ is compatible with~QM~results in the proper regime, we shall focus on the most challenging scenario, where the kinetic energy
\begin{equation}
    K
    \sim
    \frac{1}{2 m} (\langle \mathbf{p}^2 \rangle + \langle \Bar{\mathbf{p}}^2 \rangle)
    =
    \frac{\lambdabar}{2 \hbar} (\langle \mathbf{p}^2 \rangle + \langle \Bar{\mathbf{p}}^2 \rangle)
    \label{K3}
\end{equation}
can be neglected in comparison to
\begin{equation}
    |U|
    \sim
    \frac{m^2}{\sqrt{\langle |\mathbf{r} - \Bar{\mathbf{r}}|_{L_\text{C}}^2 \rangle}}
    =
    \frac{\hbar^2}{\lambdabar^2 \sqrt{\langle |\mathbf{r} - \Bar{\mathbf{r}}|_{L_\text{C}}^2 \rangle}}.
    \label{U4}
\end{equation}
Here, $\langle \ldots \rangle$ denotes mean values.

\begin{figure}[t]
   \includegraphics[width = 80mm]{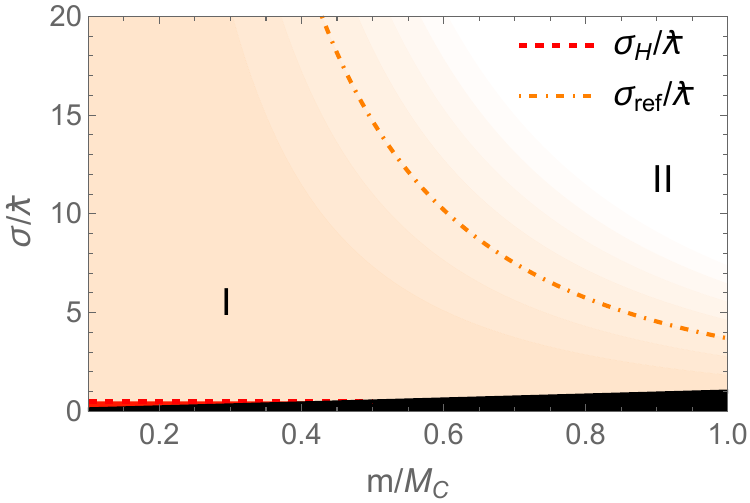}
    \caption{Region~II is the allowed parameter subspace. The dark region was excluded since it belongs to the realm of the~QST (dark region in Fig.~\ref{fig_manifesto}). The region below the dashed line ($\sigma / \lambdabar < 1 / 2$) must also be removed because it does not obey the minimum possible value for~$\sigma$ according to the Heisenberg uncertainty principle. Finally, region~I must be dismissed because it does not satisfy~$K / |U| \ll 1$ at~$t = 0$. (This plot adopts $L_\text{C} = L_\text{P}$.)}
    \label{fig_sigma}
\end{figure}

We begin checking the conditions on~$\lambdabar$ and~$\sigma$ to comply with~$K / |U| \ll 1$ at~$t = 0$. For this purpose, let us use Eq.~\eqref{W1} to calculate the following mean values in the state~$\Psi(\mathbf{r}, \Bar{\mathbf{r}}, 0)$:
\begin{equation}
    \left.\langle \mathbf{p}^2 \rangle\right|_{t = 0}
    =
    \left.\langle \Bar{\mathbf{p}}^2 \rangle\right|_{t = 0}
    =
    \frac{3 \hbar^2}{4 \sigma^2},
\end{equation}
\begin{equation}
    \left.\langle |\mathbf{r} - \Bar{\mathbf{r}}|_{L_\text{C}}^2 \rangle\right|_{t = 0}
    =
    6 \sigma^2 + L_\text{C}^2.
\end{equation}
Now, it is straightforward to use Eqs.~\eqref{K3} and~\eqref{U4} to verify that~$K / |U| \ll 1$ at~$t = 0$ provided~$\lambdabar$ and~$\sigma$ satisfy
\begin{equation}
    \left.3 \lambdabar^3 \sqrt{6 \sigma^2 + L_\text{C}^2} \right/ (4 \hbar \sigma^2)
    \ll
    1.
    \label{R1}
\end{equation}
In this case, we can neglect the kinetic term and approximate Eq.~\eqref{SE3} (at least initially) by
\begin{equation}
    i \hbar \; \partial_t \Psi(\mathbf{r}, \Bar{\mathbf{r}}, t)
    =
    - \frac{\hbar^2}{\lambdabar^2 |\mathbf{r} - \Bar{\mathbf{r}}|_{L_\text{C}}} \Psi(\mathbf{r}, \Bar{\mathbf{r}}, t).
    \label{SE4}
\end{equation}
Figure~\ref{fig_sigma} depicts the~$m$ and~$\sigma$ domain, given by Eq.~\eqref{R1}, for which Eq.~\eqref{SE4} holds. Here, $M_\text{C} = \hbar / L_\text{C} \sim M_\text{P}$ defines the Heisenberg cut --- see Fig.~\ref{fig_manifesto}. Since we explore the Heisenberg cut from the~QM~side, we are restricted to~$m / M_\text{C} < 1$. We must also exclude the dark region from the parameter space (corresponding to the dark region in Fig.~\ref{fig_manifesto}) for not complying with~$\sigma / L_\text{C} > 1$, demanded by the model. Next, we exclude region~$\sigma / \lambdabar < \sigma_\text{H} / \lambdabar \equiv 1 / 2$ (hatched region below the dashed line) for not complying with the minimum possible value for~$\sigma$ according to the Heisenberg uncertainty principle. Finally, the condition~$K / |U| \ll 1$ at~$t = 0$, given by Eq.~\eqref{R1}, imposes an extra restriction on the parameter space. Accordingly, we dismiss region~I. For the sake of reference, the dot-dashed line is defined by~$\sigma / \lambdabar = \sigma_\text{ref} / \lambdabar$, where
\begin{equation}
    \frac{\sigma_\text{ref}}{\lambdabar}
    \equiv
    \frac{3\sqrt{3}}{2}
    \frac{L_\text{C}^2 / \hbar}{(m / M_\text{C})^2} \sqrt{1 + \sqrt{1 + \frac{4 (m / M_\text{C})^6}{81 (L_\text{C}^2 / \hbar)^2}}},
    \label{B1}
\end{equation}
below which~$K / |U| > 1 / 2$ at~$t = 0$. Here, we recall that~$c = G = 1$ and~$L_\text{C}^2 / \hbar \sim L_\text{P}^2 / \hbar = 1$.

Now, let us examine for how long we can use Eq.~\eqref{SE4} to evolve~$\Psi(\mathbf{r}, \Bar{\mathbf{r}}, t)$. For this purpose, we must determine if there is an instant of time after which the kinetic term~$K$ can no longer be disregarded with respect to~$|U|$. The normalized solution of Eq.~\eqref{SE4} with initial condition~$\Psi(\mathbf{r}, \Bar{\mathbf{r}}, 0) = \psi(\mathbf{r}, 0) \psi(\Bar{\mathbf{r}}, 0)$ fixed by Eq.~\eqref{W1} is
\begin{equation}
    \Psi(\mathbf{r}, \Bar{\mathbf{r}}, t)
    =
    \frac{1}{(2 \pi \sigma^2)^{3 / 2}} 
    \exp{\left[\frac{i \hbar t}{\lambdabar^2 |\mathbf{r} - \Bar{\mathbf{r}}|_{L_\text{C}}} - \frac{\mathbf{r}^2 + \Bar{\mathbf{r}}^2}{4 \sigma^2}\right]}.
    \label{S1}
\end{equation}
In this case (see App.~\ref{appendix} for details),
\begin{equation}
    \left.\langle \mathbf{p}^2 \rangle\right|_t
    =
    \left.\langle \Bar{\mathbf{p}}^2 \rangle\right|_t
    =
    \frac{3 \hbar^2}{4 \sigma^2} 
    \left[1 + \frac{\hbar^2 A(\Tilde{\sigma})t^2}{96 \lambdabar^4 \Tilde{\sigma}^5 L_\text{C}^2}\right],
    \label{MV1}
\end{equation}
\begin{equation}
    \left.\langle |\mathbf{r} - \Bar{\mathbf{r}}|_{L_\text{C}}^2 \rangle\right|_t
    =
    6 \sigma^2 + L_\text{C}^2.
    \label{MV2}
\end{equation}
Here, $\Tilde{\sigma} \equiv \sigma / L_\text{C}$,
\begin{eqnarray}
    A(\Tilde{\sigma})
    &\equiv& 
    \sqrt{\pi} (1 + 12 \Tilde{\sigma}^2 + 12 \Tilde{\sigma}^4) e^{1 / (2 \Tilde{\sigma})^2} 
    \nonumber \\
    &\times&
    \text{erfc}[1 / (2 \Tilde{\sigma})] - 2 \Tilde{\sigma} (1 + 10 \Tilde{\sigma}^2),
\end{eqnarray}
and
\begin{equation}
    \text{erfc}(w)
    \equiv
    1 - \frac{2}{\sqrt{\pi}} \int\limits_0^w \! d w' \; e^{- w'^2}
    \label{erfc}
\end{equation}
is the complementary error function. Note that~$\left.\langle \mathbf{p}^2 \rangle\right|_t = \left.\langle \Bar{\mathbf{p}}^2 \rangle\right|_t$ grows with time, while~$\left.\langle |\mathbf{r} - \Bar{\mathbf{r}}|_{L_\text{C}}^2 \rangle\right|_t$ remains constant. With the help of Eqs.~\eqref{K3}, \eqref{U4}, \eqref{MV1}, and~\eqref{MV2}, we calculate the instant of time~$t = t_\text{ref}$,
\begin{equation}
    t_\text{ref}
    \equiv
    L_\text{C}
    \sqrt{\frac{16 \; \Tilde{\sigma}^5 [4 \Tilde{\sigma}^2 (\lambdabar L_\text{C} / \hbar) - 6 (\lambdabar^4 / \hbar^2) \sqrt{1 + 6 \Tilde{\sigma}^2}]}{\sqrt{1 + 6 \Tilde{\sigma}^2} A(\Tilde{\sigma})}},
    \label{B2}
\end{equation}
for which~$K / |U|$ grows up to~$1 / 2$. Hence, Eq.~\eqref{SE4} will not approximate Eq.~\eqref{SE3} for~$t > t_\text{ref}$. After that, the kinetic term dominates the particle-clone interaction potential, QM takes over, and no sensitive decoherence should occur.

The particle's loss of coherence can be evaluated by examining the decrease in its purity
\begin{equation}
    \eta(t)
    \equiv
    \int \! d^3 \mathbf{r} \int \! d^3 \mathbf{r}' \; |\rho(\mathbf{r}, \mathbf{r}', t)|^2
\end{equation}
as a function of time in the interval~$(0, t_\text{ref})$, where
\begin{equation}
    \rho(\mathbf{r}, \mathbf{r}', t)
    \equiv
    \int \! d^3 \Bar{\mathbf{r}} \; \Psi(\mathbf{r}, \Bar{\mathbf{r}}, t) \Psi^*(\mathbf{r}', \Bar{\mathbf{r}}, t)
\end{equation}
are elements of the particle's density matrix, obtained by tracing over the clone's degrees of freedom. We recall that this procedure is required since the clone is a mathematical artifact only introduced to model the physical transfer of information to the (inaccessible) quantum spacetime degrees of freedom.

\begin{figure}[t]
    \includegraphics[width = 80mm]{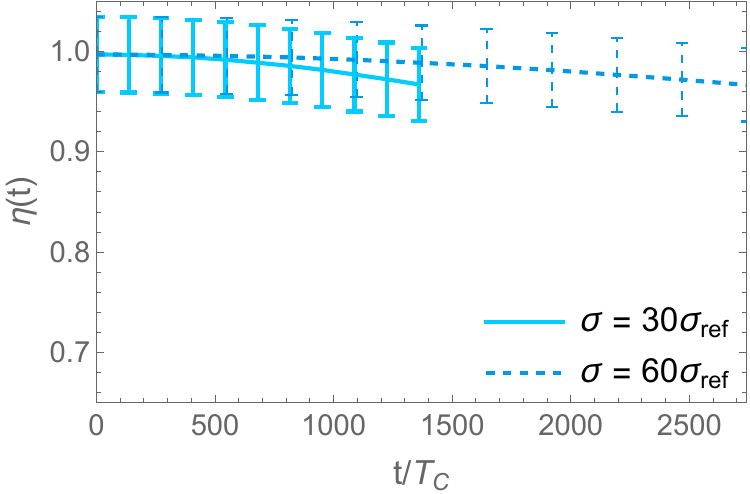}
    \caption{The time evolution of~$\eta(t)$ for a particle with mass~$m = M_\text{C} / 2$ ($\lambdabar = 2 L_\text{C}$) is exhibited for~$\sigma = 30 \; \sigma_\text{ref}$ (full line) and~$\sigma = 60 \; \sigma_\text{ref}$ (dashed line), assuming~$L_\text{C} = L_\text{P}$ ($M_\text{C} = M_\text{P}$).}
    \label{fig_eta-time}
\end{figure}

\begin{figure}[t]
    \includegraphics[width = 80mm]{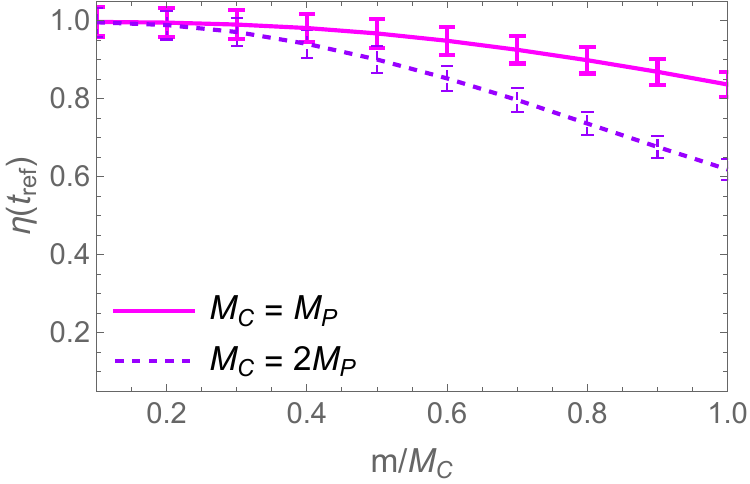}
    \caption{The purity~$\eta(t_\text{ref})$ is exhibited as a function of~$m / M_\text{C}$ ($= L_\text{C} / \lambdabar$), assuming~$L_\text{C} = L_\text{P}$ ($M_\text{C} = M_\text{P}$, solid line) and~$L_\text{C} = L_\text{P} / 2$ ($M_\text{C} = 2 M_\text{P}$, dashed line). We note that the larger the mass, the larger the decoherence. Also, larger values of~$L_\text{C}$ boost the self-decoherence for the same~$m$; e.g, fixing~$m = M_\text{P}$, we have~$\eta(t_\text{ref}) = 0.90$ for~$L_\text{C} = L_\text{P} / 2$ ($M_\text{C} = 2 M_\text{P}$), and~$\eta(t_\text{ref}) = 0.84$ for~$L_\text{C} = L_\text{P}$ ($M_\text{C} = M_\text{P}$). (We have chosen~$\sigma = 30 \; \sigma_\text{ref}$ for every value of~$m / M_\text{C}$.)}
    \label{fig_eta-mass}
\end{figure}

In Fig.~\ref{fig_eta-time} we show the time evolution of~$\eta(t)$ for a particle with mass~$m = M_\text{C} / 2$ and two values of~$\sigma$ (chosen in region~II of Fig.~\ref{fig_sigma}), where $T_\text{C} \equiv L_\text{C}$. The error bars have a numerical origin (and were evaluated by Mathematica software~\cite{rawdata}). The curves are interrupted at~$t = t_\text{ref}$. The smaller~$\sigma$ (the smaller~$t_\text{ref}$), the faster the particle decoheres. We recall, however, that, for each~$m$, there is a minimum~$\sigma$ below which the system transits to region~I of Fig.~\ref{fig_sigma}, and no significant decoherence is expected. Protons, e.g., would need to have~$\sigma > \sigma_\text{ref} \sim 10^{22}~\si{\meter}$ to be in region~II. (Still, because of their mass, $m_\text{p} / M_\text{P} \sim 10^{- 19}$, protons would virtually never self-decohere --- see forthcoming Fig.~\ref{fig_eta-mass}.)
 
\begin{figure}[b]
    \includegraphics[width = 80mm]{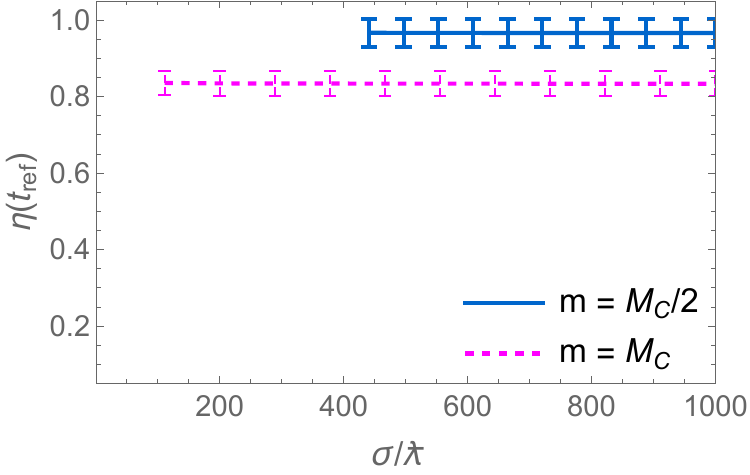}
    \caption{The purity~$\eta(t_\text{ref})$ is exhibited as a function of~$\sigma / \lambdabar$ for~$m = M_\text{C} / 2$ ($\lambdabar = 2 L_\text{C}$, full line) and~$m = M_\text{C}$ ($\lambdabar = L_\text{C}$, dashed line), assuming~$L_\text{C} = L_\text{P}$ ($M_\text{C} = M_\text{P}$). The lines have distinct domains to comply with the allowed subspace~II in Fig.~\ref{fig_sigma}. The larger the mass, the larger the decoherence.}
    \label{fig_eta-sigma}
\end{figure}

Interestingly, the purity~$\eta(t_\text{ref})$ for systems in region~II does not seem sensitive to~$\sigma$. This is made explicit in Fig.~\ref{fig_eta-sigma} where~$\eta(t_\text{ref})$ is plotted as a function of~$\sigma$ for two mass values. The lines start at different~$\sigma / \lambdabar$ to guarantee they are inside region~II. The continuation of these lines to smaller values of~$\sigma / \lambdabar$ should converge fast to~$\eta(t_\text{ref}) \to 1$ since no decoherence would be expected for systems in region~I. We see that the larger the mass~$m$, the smaller the purity~$\eta(t_\text{ref})$.

Finally, Fig.~\ref{fig_eta-mass} plots the purity~$\eta(t_\text{ref})$ as a function of~$m / M_\text{C}$ for different values of~$M_\text{C}$ and carries the information that the larger the mass, the larger the gravitational self-decoherence. Equally interesting is that, for a fixed mass~$m$, larger values of~$L_\text{C}$ (smaller values of~$M_\text{C}$) boost the self-decoherence [leading to smaller values of~$\eta(t_\text{ref})$]. For instance, fixing~$m = M_\text{P}$, we have~$\eta(t_\text{ref}) = 0.90$ for~$L_\text{C} = L_\text{P} / 2$ ($M_\text{C} = 2 M_\text{P}$), and~$\eta(t_\text{ref}) = 0.84$ for~$L_\text{C} =  L_\text{P}$ ($M_\text{C} = M_\text{P}$).

It is clear that the present proposal differs from several others in terms of the relevance of delocalization. In the~DP~model and similar ones, e.g., the larger the delocalization, the larger the metric fluctuations, and the faster the system decoheres. In the present model, Figs.~\ref{fig_eta-time} and~\ref{fig_eta-sigma} indicate that, once the system is in region~II, its delocalization will impact how fast the~CoM decoheres but not the ``amount of decoherence." The {\em amount of decoherence} is ruled by the mass~$m$, reflecting the Heisenberg cut, which separates quantum ($m \ll M_\text{P}$) and classical ($m \gg M_\text{P}$) realms.

\section{Connecting the model to cutting-edge experiments}
\label{sec_experiments}

Recent remarkable results indicate that systems composed of a seemingly arbitrarily large number of microscopic constituents can exhibit quantum coherence. Although this may challenge models where the number of quantum constituents matters to explain the classical behavior of macroscopic systems, it is not the case for the present model.

In Ref.~\cite{1997-andrews}, interference of two freely expanding Bose-Einstein condensates with~$5 \times 10^6$~Na atoms separated by~$40~\si{\micro\meter}$ has been observed. As interesting as it may be, the fringes associated with de Broglie wavelength~$\lambda = h / (m v) \approx 17~\si{\micro\meter}$ (with~$m v \sim 10^{- 3}~\si{\meter / \second}$) concern the Na atoms ($m_\text{Na} \approx 3.8 \times 10^{- 26}~\si{\kilo\gram}$), not the~CoM of the condensate, whose fringes would be a million times smaller. Hence, this experiment does not test the interference of the~CoM, which concerns the present model. Moreover, according to Fig.~\ref{fig_eta-mass}, condensates of masses~$10^{- 12}~M_\text{P}$ would virtually experience no gravitational self-decoherence.

More recently, Bild et al.~\cite{2023-bild} announced a result interpreted as a compelling case for the validity of~QM for macroscopic bodies. Briefly, they have put a coherent phonon state in a quantum superposition of two opposite-phase oscillations. As the authors state, however, they are not considering degrees of freedom of the~CoM but of perturbations in a (high-overtone-bulk-acoustic-wave) resonator. Thus, although the crystal lattice has~$10^{17}$ atoms and a mass of~$16~\si{\micro\gram} \sim M_\text{P}$, the thinking that these phonons would delocalize the~CoM~of the system is disputable. It is hard to say what would happen if they had energies of the order of~$M_\text{P}$ ($10^{19}~\si{\giga\electronvolt}$), but they do not; their energies are of the order of~$\hbar \times 6~\si{\giga\hertz} \approx 4 \times 10^{- 15}~\si{\giga\electronvolt}$. Although phonons are massless and, thus, beyond the scope of this paper, our result that nonrelativistic particles with masses~$m \ll M_\text{P}$ should obey usual~QM suggests that phonons should be well-described by~QM {\em provided} they have energies~$E_\text{phonon} \ll M_\text{P}$ as measured in the rest frame of the background medium (being in harmony with Bild et al.).

The present model makes us believe that any small perturbations (i.e., far from the Planck scale), including collective perturbations of an arbitrary number of quantum constituents, should satisfy usual~QM~\cite{2013-purdy, 2021-whittle}, {\em yet propagating in a medium with mass~$m \gg M_\text{P}$ with~CoM ruled by classical physics.} In this vein, massless spin-2 perturbations, e.g., of black holes, would be well-described by usual quantum field theory in curved spacetimes. Yet, this would not imply black holes (themselves) could be superposed as if they were quantum systems with~$m \ll M_\text{P}$~\cite{2022-foo}. From the present perspective, black holes should be classical structures dwelling on the right-hand side of the Heisenberg cut (see Fig.~\ref{fig_manifesto}). To be ``quantum," the black-hole masses should satisfy~$m \ll M_\text{P}$, in which case their reduced Compton wavelengths would leak outside the horizon defined by the Schwarzschild radius: $\lambdabar_\text{BH} \gg R_\text{Sc}$, and it is hard to see how such structures could be considered black holes. Hopefully, similar considerations may mitigate problems concerning the quantum-to-classical transition of fluctuations created in the early Universe as reported, e.g., in Refs.~\cite{2011-sudarsky, 2021-berjon}.

Decades of diligent search suggest that Einstein's equations and their solutions (including black holes) may be too far from the~QST to be a good starting point for quantum gravity. From this point of view, the semiclassical Einstein equations (and the Schr{\" o}dinger-Newton equation linked to them in the nonrelativistic limit) finish up disfavored~\cite{2023-silva}.

This leads us to wonder how the present proposal can be tested (see Ref.~\cite{2016-pfister} for a universal strategy to test gravitational decoherence). Inspired by Ref.~\cite{2005-lindner}, one could imagine scattering quantum particles by a massive enough system superposed in space. According to the present scenario, the system would gravitationally self-decohere, and the resulting interference fringes would correspond to the ones of a combination of two single-slit experiments instead of the usual double-slit experiment. This strategy is as conceptually simple as experimentally challenging. Even using massive systems such as macromolecules, they would account for a mass up to~$m \sim 10^4~\si{\dalton} \sim 10^{- 15}~M_\text{P}$. Cutting-edge technology can keep the coherence of such systems for no more than some milliseconds at the moment~\cite{2019-fein}. This is far from what would be necessary to test our proposal today.

Next, we discuss a strategy to test our model based on a double Stern-Gerlach experiment followed by a spin measurement and compare the results with the ones obtained when a similar strategy is used to test the Schr{\" o}dinger-Newton equation~\cite{2024-aguiar, 2024-grossardt}.

\section{Gravitational self-decoherence in a Stern-Gerlach-like experiment}
\label{sec_sg}

Let us consider a particle with reduced Compton wavelength~$\lambdabar$ and spin~$s = 1 / 2$ initially described by a Gaussian state
\begin{eqnarray}
    | \psi(0) \rangle
    &=&
    \sum_{\zeta = \uparrow, \downarrow} \psi_\zeta(\mathbf{r}, 0) | \zeta \rangle
    \nonumber \\   
    &=&
    \psi_\uparrow(\mathbf{r}, 0) | \uparrow \rangle + \psi_\downarrow(\mathbf{r}, 0) | \downarrow \rangle,
    \label{SV1}
\end{eqnarray}
where
\begin{equation}
    \psi_\zeta(\mathbf{r}, 0)
    \equiv
    \frac{1}{\sqrt{2} (2 \pi \sigma^2)^{3/4}} \exp{\left(- \frac{\mathbf{r}^2}{4 \sigma^2}\right)}.
    \label{W2}
\end{equation}
Here, $\{| \uparrow \rangle, | \downarrow \rangle\}$ is the eigenstate basis of the $z$-axis spin operator~$\Hat{S}_z$:
\begin{equation}
    \Hat{S}_z | \uparrow \rangle
    \equiv
    (+ \hbar / 2) | \uparrow \rangle,
    \quad
    \Hat{S}_z | \downarrow \rangle
    \equiv
    (- \hbar / 2) | \downarrow \rangle.
\end{equation}
(Note that the wavefunction~\eqref{W2} does not depend on~$\zeta$, and, therefore, the particle's state~\eqref{SV1} --- which precedes the Stern-Gerlach-like experiment --- does not exhibit entanglement between position and spin degrees of freedom.) Now, we take the tensor product of state~\eqref{SV1} with a copy of it (representing the clone) to build the initial particle-clone (normalized) state:
\begin{eqnarray}
    | \Psi(0) \rangle
    &\equiv&
    \sum_{\zeta = \uparrow, \downarrow} \psi_\zeta(\mathbf{r}, 0) | \zeta \rangle \otimes \sum_{\Bar{\zeta} = \uparrow, \downarrow} \psi_{\Bar{\zeta}}(\Bar{\mathbf{r}}, 0) | \Bar{\zeta} \rangle
    \nonumber \\
    &=&
    \sum_{\zeta, \Bar{\zeta} = \uparrow, \downarrow} \Psi_{\zeta \Bar{\zeta}}(\mathbf{r}, \Bar{\mathbf{r}}, 0) | \zeta, \Bar{\zeta} \rangle,
    \label{SV2}
\end{eqnarray}
where
\begin{equation}
    \Psi_{\zeta \Bar{\zeta}}(\mathbf{r}, \Bar{\mathbf{r}}, 0)
    \equiv
    \psi_\zeta(\mathbf{r}, 0) \psi_{\Bar{\zeta}}(\Bar{\mathbf{r}}, 0).
\end{equation}

\begin{figure}[t]
    \includegraphics[width=\columnwidth]{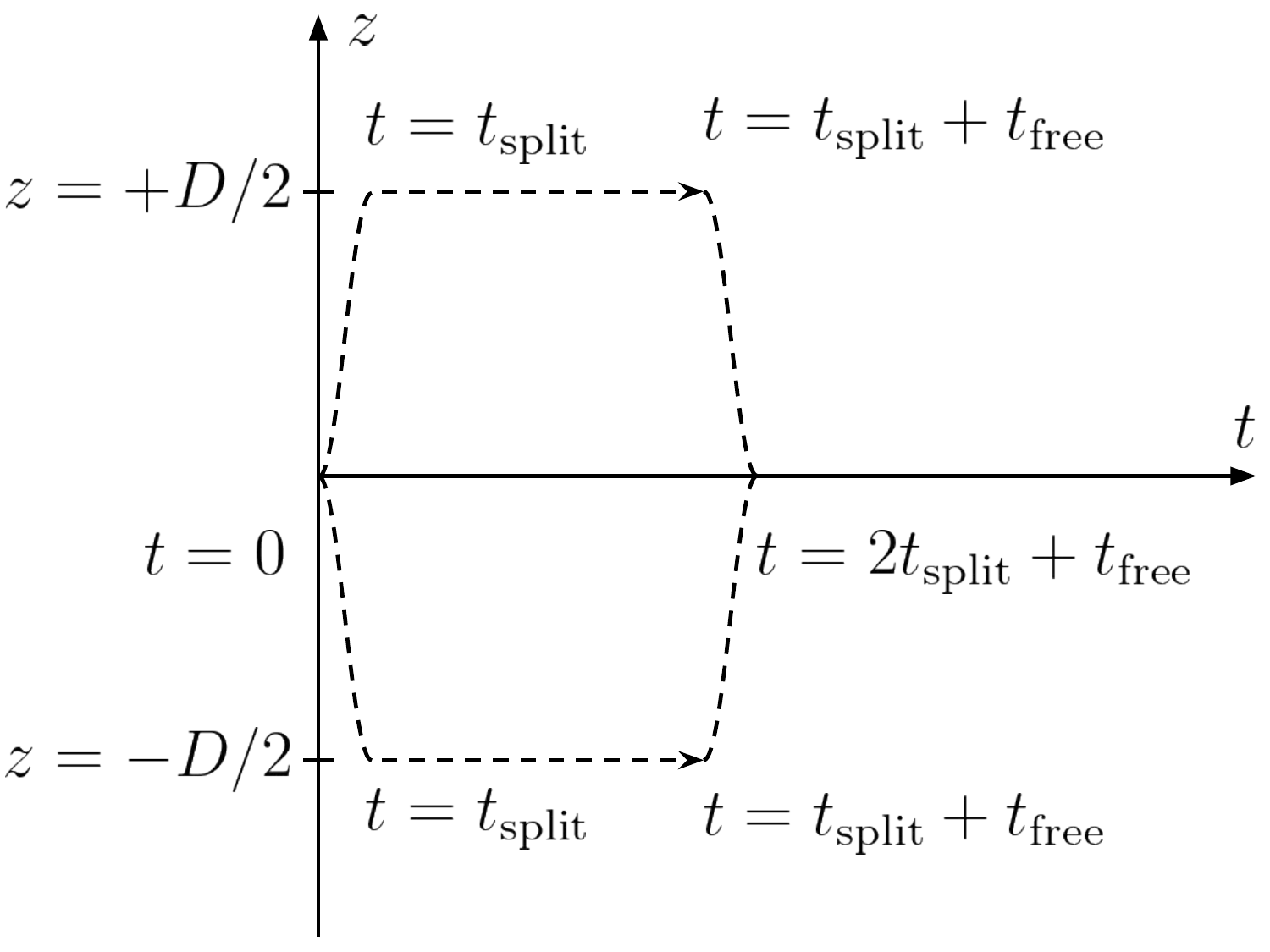}
    \caption{The system starts in state~$| \Psi(0) \rangle$, transits to~$| \Psi(t_\text{split}) \rangle$ as it is (Stern-Gerlach) split, evolves to~$| \Psi(t_\text{split} + t_\text{free}) \rangle$, and ends up as~$| \Psi(2 t_\text{split} + t_\text{free}) \rangle$ after the Stern-Gerlach split is reversed.}
    \label{fig_diagram}
\end{figure}

Next, the initial state~\eqref{SV2} is (Stern-Gerlach) split along the $z$~axis, so the combined particle-clone system gets spatially superposed by a distance~$D$ (see Fig.~\ref{fig_diagram}) as represented by the state
\begin{equation}
    | \Psi(t_\text{split}) \rangle
    =
    \sum_{\zeta, \Bar{\zeta} = \uparrow, \downarrow} \Psi_{\zeta \Bar{\zeta}}(\mathbf{r}, \Bar{\mathbf{r}}, t_\text{split}) | \zeta, \Bar{\zeta} \rangle,
    \label{SV3}
\end{equation}
where
\begin{equation}
    \Psi_{\zeta \Bar{\zeta}}(\mathbf{r}, \Bar{\mathbf{r}}, t_\text{split})
    =
    \Psi_{\zeta \Bar{\zeta}}(\mathbf{r} - \mathbf{R}_\zeta, \Bar{\mathbf{r}} - \mathbf{R}_{\Bar{\zeta}}, 0),
    \label{W3}
\end{equation}
\begin{equation}
    \mathbf{R}_\uparrow \equiv (0, 0, + D / 2),
    \quad
    \mathbf{R}_\downarrow \equiv (0, 0, - D / 2),
\end{equation}
and we have disregarded the interaction potential~\eqref{U2} during the tiny splitting time interval~$t_\text{split}$. Note that Eq.~\eqref{SV3} is particle-clone symmetric, as it should be.

Next, the system is assumed to evolve free of {\em external} influences, $V(\mathbf{r}, t) = V(\Bar{\mathbf{r}}, t) = 0$, in the time interval~$(t_\text{split}, t_\text{split} + t_\text{free})$, along which~$\Psi_{\zeta \Bar{\zeta}}(\mathbf{r}, \Bar{\mathbf{r}}, t)$ will satisfy Eq.~\eqref{SE3} with~$\Psi(\mathbf{r}, \Bar{\mathbf{r}}, t) \to \Psi_{\zeta \Bar{\zeta}}(\mathbf{r}, \Bar{\mathbf{r}}, t)$, namely,
\begin{eqnarray}
    i \hbar \; \partial_t \Psi_{\zeta \Bar{\zeta}}(\mathbf{r}, \Bar{\mathbf{r}}, t)
    &=&
    - \frac{\hbar \lambdabar}{2} (\nabla_\mathbf{r}^2 + \nabla_{\Bar{\mathbf{r}}}^2) \Psi_{\zeta \Bar{\zeta}}(\mathbf{r}, \Bar{\mathbf{r}}, t)
    \nonumber \\
    &{}&
    - \frac{\hbar^2}{\lambdabar^2 |\mathbf{r} - \Bar{\mathbf{r}}|_{L_\text{C}}} \Psi_{\zeta \Bar{\zeta}}(\mathbf{r}, \Bar{\mathbf{r}}, t).
    \label{SE5}
\end{eqnarray}
To address this evolution, we shall compare [see Eqs.~\eqref{K3} and~\eqref{U4}]
\begin{equation}
    K_{\zeta \Bar{\zeta}}
    \sim
    \frac{1}{2 m} \left(\langle \mathbf{p}^2 \rangle_{\zeta \Bar{\zeta}} + \langle \Bar{\mathbf{p}}^2 \rangle_{\zeta \Bar{\zeta}}\right)
    =
    \frac{\lambdabar}{2 \hbar} \left(\langle \mathbf{p}^2 \rangle_{\zeta \Bar{\zeta}} + \langle \Bar{\mathbf{p}}^2 \rangle_{\zeta \Bar{\zeta}}\right)
    \label{K4}
\end{equation}
and
\begin{equation}
    |U_{\zeta \Bar{\zeta}}|
    \sim
    \frac{m^2}{\sqrt{\langle |\mathbf{r} - \Bar{\mathbf{r}}|_{L_\text{C}}^2 \rangle_{\zeta \Bar{\zeta}}}}
    =
    \frac{\hbar^2}{\lambdabar^2 \sqrt{\langle |\mathbf{r} - \Bar{\mathbf{r}}|_{L_\text{C}}^2 \rangle_{\zeta \Bar{\zeta}}}},
    \label{U5}
\end{equation}
where the mean values~$\langle \ldots \rangle_{\zeta \Bar{\zeta}}$ are calculated here with respect to the corresponding wavefunction~$\Psi_{\zeta \Bar{\zeta}}(\mathbf{r}, \Bar{\mathbf{r}}, t)$ (after proper normalization). Let us analyze separately the evolution of~$\Psi_{\zeta \Bar{\zeta}}(\mathbf{r}, \Bar{\mathbf{r}}, t)$ for the cases ($a$)~$\zeta = \Bar{\zeta}$ and ($b$)~$\zeta \neq \Bar{\zeta}$.

\paragraph{$\Psi_{\zeta \Bar{\zeta}}(\mathbf{r}, \Bar{\mathbf{r}}, t)$ {\rm with}~$\zeta = \Bar{\zeta}$ {\rm :}}\;

In this case,
\begin{equation}
   \left.\langle \mathbf{p}^2 \rangle_{\zeta \Bar{\zeta}}\right|_{t = t_\text{split}}
    =
   \left.\langle \Bar{\mathbf{p}}^2 \rangle_{\zeta \Bar{\zeta}}\right|_{t = t_\text{split}}
    = \frac{3 \hbar^2}{4 \sigma^2},
\end{equation}
\begin{equation}
    \left.\langle |\mathbf{r} - \Bar{\mathbf{r}}|_{L_\text{C}}^2 \rangle_{\zeta \Bar{\zeta}}\right|_{t = t_\text{split}}
    =
    6 \sigma^2 + L_\text{C}^2,
\end{equation}
from which it is straightforward to use Eqs.~\eqref{K4} and~\eqref{U5} to verify that~$K_{\zeta \Bar{\zeta}} / |U|_{\zeta \Bar{\zeta}} \ll 1$ at~$t = t_\text{split}$, provided~$\lambdabar$ and~$\sigma$ satisfy the condition~\eqref{R1}. (Recall that the choice of~$\lambdabar$ and~$\sigma$ in region~II of Fig.~\ref{fig_sigma} automatically complies with Eq.~\eqref{R1}.) In this case, we can neglect the kinetic term and approximate Eq.~\eqref{SE5} by [see Eq.~\eqref{SE4}]
\begin{equation}
    i \hbar \; \partial_t \Psi_{\zeta  \Bar{\zeta}}(\mathbf{r}, \Bar{\mathbf{r}}, t)
    =
    - \frac{\hbar^2}{\lambdabar^2 |\mathbf{r} - \Bar{\mathbf{r}}|_{L_\text{C}}} \Psi_{\zeta \Bar{\zeta}}(\mathbf{r}, \Bar{\mathbf{r}}, t),
    \label{SE6}
\end{equation}
at least initially.

Next, we must investigate for how long we can use Eq.~\eqref{SE6} to evolve~$\Psi_{\zeta \Bar{\zeta}}(\mathbf{r}, \Bar{\mathbf{r}}, t)$. For this purpose, we must determine if there is an instant of time after which the kinetic term~$K_{\zeta \Bar{\zeta}}$ can no longer be disregarded with respect to~$|U|_{\zeta \Bar{\zeta}}$. The normalized solution of Eq.~\eqref{SE6} with initial condition~\eqref{W3} is
\begin{eqnarray}
    \Psi_{\zeta \Bar{\zeta}}(\mathbf{r}, \Bar{\mathbf{r}}, t)
    &=&
    \frac{1}{2 (2 \pi \sigma^2)^{3 / 2}} \exp{\left[\frac{i \hbar \; (t - t_\text{split})}{\lambdabar^2 |\mathbf{r} - \Bar{\mathbf{r}}|_{L_\text{C}}}\right]}
    \nonumber \\
    &\times&
    \exp{\left[\frac{- |\mathbf{r} - \mathbf{R}_\zeta|^2 - |\Bar{\mathbf{r}} - \mathbf{R}_{\Bar{\zeta}}|^2}{4 \sigma^2}\right]}.
    \label{S2}
\end{eqnarray}
In this case,
\begin{eqnarray}
    \left.\langle \mathbf{p}^2 \rangle_{\zeta \Bar{\zeta}}\right|_t
    &=&
    \left.\langle \Bar{\mathbf{p}}^2 \rangle_{\zeta \Bar{\zeta}}\right|_t
    \nonumber \\
    &=&
    \frac{3 \hbar^2}{4 \sigma^2} 
    \left[1 + \frac{\hbar^2 A(\Tilde{\sigma}) (t - t_\text{split})^2}{96 \lambdabar^4 \Tilde{\sigma}^5 L_\text{C}^2}\right], 
    \label{MV3}
\end{eqnarray}
\begin{equation}
    \left.\langle |\mathbf{r} - \Bar{\mathbf{r}}|_{L_\text{C}}^2 \rangle_{\zeta \Bar{\zeta}}\right|_t
    =
    6 \sigma^2 + L_\text{C}^2.
    \label{MV4}
\end{equation}
Following the discussion below Eq.~\eqref{erfc}, we note that~$\left.\langle \mathbf{p}^2 \rangle_{\zeta \Bar{\zeta}}\right|_t = \left.\langle \Bar{\mathbf{p}}^2 \rangle_{\zeta \Bar{\zeta}} \right|_t$ grows with time, while~$\left.\langle |\mathbf{r} - \Bar{\mathbf{r}}|_{L_\text{C}}^2 \rangle_{\zeta \Bar{\zeta}}\right|_t$ remains constant. Then, by using Eqs.~\eqref{K4}, \eqref{U5}, \eqref{MV3}, and~\eqref{MV4}, we obtain that~$K_{\zeta \Bar{\zeta}} / |U|_{\zeta \Bar{\zeta}} < 1 / 2$ in the interval
\begin{equation}
    (t_\text{split}, t_\text{split} + t_\text{ref}), 
\end{equation}
where~$t_\text{ref}$ is defined in Eq.~\eqref{B2}.


\paragraph{$\Psi_{\zeta \Bar{\zeta}}(\mathbf{r}, \Bar{\mathbf{r}}, t)$ {\rm with}~$\zeta \neq \Bar{\zeta}$ {\rm :}}\;

Following the same steps as for~$\zeta = \Bar{\zeta}$, we have here
\begin{equation}
    \left.\langle \mathbf{p}^2 \rangle_{\zeta \Bar{\zeta}}\right|_{t = t_\text{split}}
    =
    \left.\langle \Bar{\mathbf{p}}^2 \rangle_{\zeta \Bar{\zeta}}\right|_{t = t_\text{split}}
    =
    \frac{3 \hbar^2}{4 \sigma^2},
    \label{MV5}
\end{equation}
\begin{equation}
    \left.\langle |\mathbf{r} - \Bar{\mathbf{r}}|_{L_\text{C}}^2 \rangle_{\zeta \Bar{\zeta}}\right|_{t = t_\text{split}}
    =
    6 \sigma^2 + D^2 + L_\text{C}^2.
    \label{MV6}
\end{equation}
Then, for large enough~$D$,
\begin{equation}
    D
    \gg
    \sigma
    \gtrsim
    \lambdabar
    \gtrsim
    L_\text{C},
    \label{R2}
\end{equation}
we obtain that~$K_{\zeta \Bar{\zeta}} / |U|_{\zeta \Bar{\zeta}} \gg 1$ at~$t = t_\text{split}$ and, now, it is the particle-clone interaction potential that can be neglected. Then, in opposition to the previous case, here, Eq.~\eqref{SE5} can be initially approximated by
\begin{equation}
    i \hbar \; \partial_t \Psi_{\zeta \Bar{\zeta}}(\mathbf{r}, \Bar{\mathbf{r}}, t)
    =
    - \frac{\hbar \lambdabar}{2} (\nabla_\mathbf{r}^2 + \nabla_{\Bar{\mathbf{r}}}^2) \Psi_{\zeta \Bar{\zeta}}(\mathbf{r}, \Bar{\mathbf{r}}, t).
    \label{SE7}
\end{equation}

The normalized solution of Eq.~\eqref{SE7} with initial condition~\eqref{W3} is
\begin{eqnarray}
    \Psi_{\zeta \Bar{\zeta}}(\mathbf{r}, \Bar{\mathbf{r}}, t)
    &=&
    \frac{[1 + i \lambdabar (t - t_\text{split}) / (2 \sigma^2)]^{-3}}{2 (2 \pi \sigma^2)^{3 / 2}} 
    \nonumber \\
    &\times&
    \exp{\left\{\frac{- |\mathbf{r} - \mathbf{R}_\zeta |^2 - |\Bar{\mathbf{r}} - \mathbf{R}_{\Bar{\zeta}}|^2}{4 \sigma^2 [1 + i \lambdabar (t - t_\text{split}) / (2 \sigma^2)]}\right\}}.
    \nonumber \\
    &{}&
    \label{S3}
\end{eqnarray}
Here, let us assume that the system is evolved up to~$t_\text{split} + t_\text{free}$, where
\begin{equation}
    t_\text{free}
    \leq
    t_\text{ref}
    \ll
    2 \sigma^2 / \lambdabar. 
    \label{R3}
\end{equation}
In the time interval
\begin{equation}
    (t_\text{split}, t_\text{split} + t_\text{free}),
    \label{I}
\end{equation}
Eq.~\eqref{S3} simplifies as
\begin{eqnarray}
    \Psi_{\zeta \Bar{\zeta}}(\mathbf{r}, \Bar{\mathbf{r}}, t)
    &=&
    \frac{1}{2 (2 \pi \sigma^2)^{3 / 2}}
    \nonumber \\
    &\times&
    \exp{\left[\frac{- |\mathbf{r} - \mathbf{R}_\zeta|^2 - |\Bar{\mathbf{r}} - \mathbf{R}_{\Bar{\zeta}}|^2}{4 \sigma^2}\right]},
    \label{S4}
\end{eqnarray}
and~$\left.\langle \mathbf{p}^2 \rangle_{\zeta \Bar{\zeta}}\right|_t = \left.\langle \mathbf{\Bar{p}}^2 \rangle_{\zeta \Bar{\zeta}}\right|_t$ and~$\left.\langle |\mathbf{r} - \Bar{\mathbf{r}}|_{L_\text{C}}^2 \rangle_{\zeta \Bar{\zeta}}\right|_t$ are given as in the right-hand side of Eqs.~\eqref{MV5} and~\eqref{MV6}, respectively. From this, we see that $K_{\zeta \Bar{\zeta}} / |U|_{\zeta \Bar{\zeta}} \gg 1$ and Eq.~\eqref{SE7} can be used in the interval~\eqref{I}.

Thus, by freely evolving the system in the time interval~$(t_\text{split}, t_\text{split} + t_\text{free})$ [with~$t_\text{free}$ given by Eq.~\eqref{R3}], we are authorized to use Eqs.~\eqref{S2} and~\eqref{S4} to describe~$\Psi_{\zeta \Bar{\zeta}} (\mathbf{r}, \Bar{\mathbf{r}}, t)$ for~$\zeta = \Bar{\zeta}$ and~$\zeta \neq \Bar{\zeta}$, respectively.

Next, we reverse the Stern-Gerlach split (neglecting the interaction potential~\eqref{U2} as in the splitting). The resulting state is
\begin{equation}
    | \Psi(2 t_\text{split} + t_\text{free}) \rangle
    =
    \sum_{\zeta, \Bar{\zeta} = \uparrow, \downarrow} \Psi_{\zeta \Bar{\zeta}}(\mathbf{r}, \Bar{\mathbf{r}}, 2 t_\text{split} + t_\text{free}) | \zeta, \Bar{\zeta} \rangle,
    \label{F}
\end{equation}
where
\begin{equation}
    \Psi_{\zeta \Bar{\zeta}}(\mathbf{r}, \Bar{\mathbf{r}}, 2 t_\text{split} + t_\text{free})
    =
    \Psi_{\zeta \Bar{\zeta}}(\mathbf{r} + \mathbf{R}_\zeta, \Bar{\mathbf{r}} + \mathbf{R}_{\Bar{\zeta}}, t_\text{split} + t_\text{free})
\end{equation}
was obtained by turning over the splitting [see Eq.~\eqref{W3}]. Note that Eq.~\eqref{F} is symmetric concerning particle-and-clone exchange, as it should be.

As discussed above, the clone is a mathematical artifact introduced to model the physical transfer of information from the particle to the (inaccessible) quantum spacetime degrees of freedom. Thus, the clone's degrees of freedom must be traced over eventually. Proceeding accordingly, we end up with the particle's density matrix, which can be represented by the elements
\begin{equation}
    \rho_{\zeta \zeta'}(\mathbf{r}, \mathbf{r}', t)
    \equiv
    \sum_{\Bar{\zeta} = \uparrow, \downarrow} \int \! d^3 \Bar{\mathbf{r}} \; \Psi_{\zeta \Bar{\zeta}}(\mathbf{r}, \Bar{\mathbf{r}}, t) \Psi_{\zeta' \Bar{\zeta}}^*(\mathbf{r}', \Bar{\mathbf{r}}, t).
    \label{E}
\end{equation}

\begin{figure}[t]
    \includegraphics[width = 80mm]{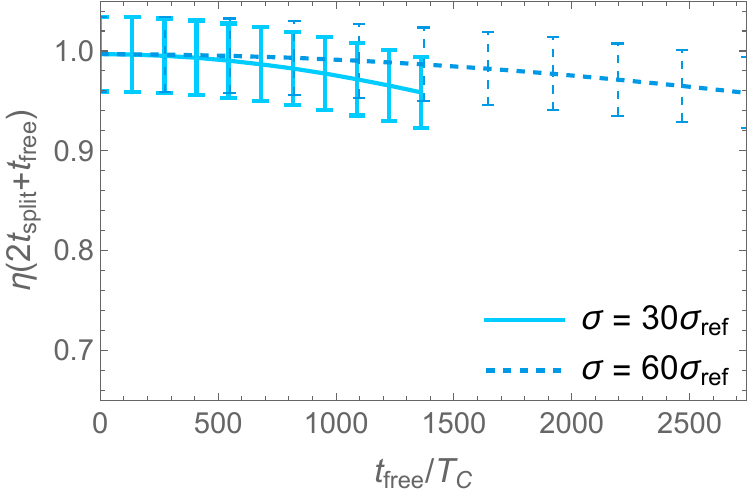}
    \caption{The purity~$\eta(2 t_\text{split} + t_\text{free})$ for a particle with mass~$m = M_\text{C} / 2$ ($\lambdabar = 2 L_\text{C}$) is exhibited as a function of~$t_\text{free}$ for~$\sigma = 30 \; \sigma_\text{ref}$ (full line) and~$\sigma = 60 \; \sigma_\text{ref}$ (dashed line). Here, $D = 100 \; \sigma^2 / L_\text{C}$ and~$L_\text{C} = L_\text{P}$ ($M_\text{C} = M_\text{P}$).}
    \label{fig_mu-time}
\end{figure}

Let us begin discussing the purity of the particle's state:
\begin{equation}
    \eta(t)
    \equiv
    \sum_{\zeta, \zeta' = \uparrow, \downarrow} \int \! d^3 \mathbf{r} \int \! d^3 \mathbf{r}' \; |\rho_{\zeta \zeta'}(\mathbf{r}, \mathbf{r}', t)|^2.
\end{equation}
In Fig.~\ref{fig_mu-time} we plot~$\eta(2 t_\text{split} + t_\text{free})$ as a function of~$t_\text{free}$ for a particle with mass~$m = M_\text{C} / 2$ and two values of~$\sigma$ (chosen in region~II of Fig.~\ref{fig_sigma}). The curves are interrupted at~$t_\text{free} = t_\text{ref}$ [see Eq.~\eqref{R3}]. Consistent with the discussion in Sec.~\ref{sec_free}, the smaller~$\sigma$, the faster the particle decoheres. As in Fig.~\ref{fig_sigma}, for each~$m$, there is a minimum~$\sigma$ below which no significant decoherence should occur.

\begin{figure}[b]
    \includegraphics[width = 80mm]{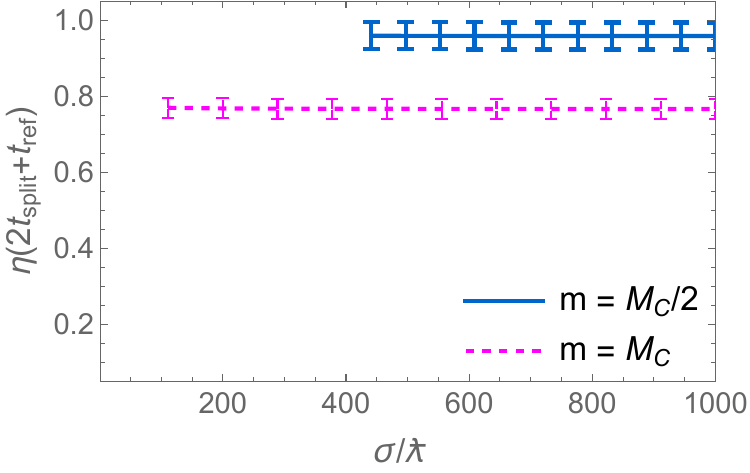}
    \caption{The purity~$\eta(2 t_\text{split} + t_\text{ref})$ is exhibited as a function of~$\sigma / \lambdabar$ for~$m = M_\text{C} / 2$ ($\lambdabar = 2 L_\text{C}$, full line) and~$m = M_\text{C}$ ($\lambdabar = L_\text{C}$, dashed line), assuming~$D = 100 \; \sigma^2 / L_\text{C}$ and~$L_\text{C} = L_\text{P}$ ($M_\text{C} = M_\text{P}$). The larger the mass, the larger the decoherence.}
    \label{fig_mu-sigma}
\end{figure}
 
Figure~\ref{fig_mu-sigma} plots the purity~$\eta(2 t_\text{split} + t_\text{ref})$ as a function of~$\sigma$ for distinct mass values. The larger the mass~$m$, the smaller the purity~$\eta(2 t_\text{split} + t_\text{ref})$. Figure~\ref{fig_mu-sigma} is similar, but not identical, to the analogous Fig.~\ref{fig_eta-sigma}.

\begin{figure}[t]
    \includegraphics[width = 80mm]{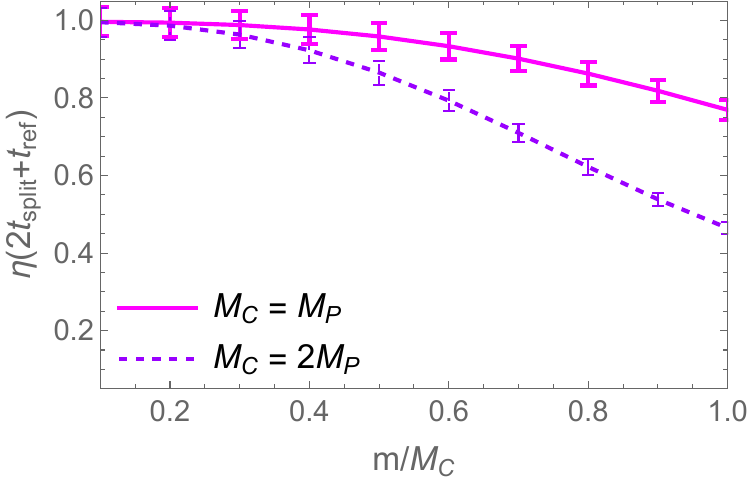}
    \caption{The purity~$\eta(2 t_\text{split} + t_\text{ref})$ is exhibited as a function of~$m / M_\text{C}$ ($= L_\text{C} / \lambdabar$), assuming~$L_\text{C} = L_\text{P}$ ($M_\text{C} = M_\text{P}$, solid line) and~$L_\text{C} = L_\text{P} / 2$ ($M_\text{C} = 2 M_\text{P}$, dashed line). The larger the mass, the larger the decoherence. Also, larger values of~$L_\text{C}$ boost the self-decoherence for the same~$m$; e.g, fixing~$m = M_\text{P}$, we have~$\eta(2 t_\text{split} + t_\text{ref}) = 0.87$ for~$L_\text{C} = L_\text{P} / 2$ ($M_\text{C} = 2 M_\text{P}$), and~$\eta(2 t_\text{split} + t_\text{ref}) = 0.77$ for~$L_\text{C} = L_\text{P}$ ($M_\text{C} = M_\text{P}$). (We have chosen~$\sigma = 30 \; \sigma_\text{ref}$ and~$D = 100 \; \sigma^2 / L_\text{C}$ for every value of~$m / M_\text{C}$.)}
    \label{fig_mu-mass}
\end{figure}

Next, Fig.~\ref{fig_mu-mass} plots the purity~$\eta(2 t_\text{split} + t_\text{ref})$ as a function of~$m / M_\text{C}$ for different values of~$M_\text{C}$. As expected from the previous figure (see also discussion in Sec.~\ref{sec_free}), the larger the mass, the larger the gravitational self-decoherence. By comparing Figs.~\ref{fig_mu-mass} and~\ref{fig_eta-mass}, we see that the decoherence in the Stern-Gerlach case is greater than in the free-particle case. The increase in the Stern-Gerlach case is due to the presence of spin (recall that~$\Psi_{\zeta \Bar{\zeta}}(\mathbf{r}, \Bar{\mathbf{r}}, t)$ evolves differently for~$\zeta = \Bar{\zeta}$ and~$\zeta \neq \Bar{\zeta}$). Again, once mass~$m$ is fixed, larger values of~$L_\text{C}$ boost the self-decoherence. (For instance, fixing~$m = M_\text{P}$, we have~$\eta(2 t_\text{split} + t_\text{ref}) = 0.87$ for~$L_\text{C} = L_\text{P} / 2$ ($M_\text{C} = 2 M_\text{P}$), and~$\eta(2 t_\text{split} + t_\text{ref}) = 0.77$ for~$L_\text{C} =  L_\text{P}$ ($M_\text{C} = M_\text{P}$).)

Now, we progress and show how the gravitational self-decoherence imprints a signal on the spin, distinguishing our model from~QM. By tracing over the position degrees of freedom from the particle's density matrix, we obtain the reduced density matrix restricted to spin:
\begin{equation}
    \Hat{\varrho}(t)
    \equiv
    \sum_{\zeta, \zeta' = \uparrow, \downarrow} \int \! d^3 \mathbf{r} \; \rho_{\zeta \zeta'}(\mathbf{r}, \mathbf{r}, t) | \zeta \rangle \langle \zeta' |,
\end{equation}
where~$\rho_{\zeta \zeta'}(\mathbf{r}, \mathbf{r}', t)$ is given in Eq.~\eqref{E}.

\begin{figure}[t]
    \includegraphics[width = 80mm]{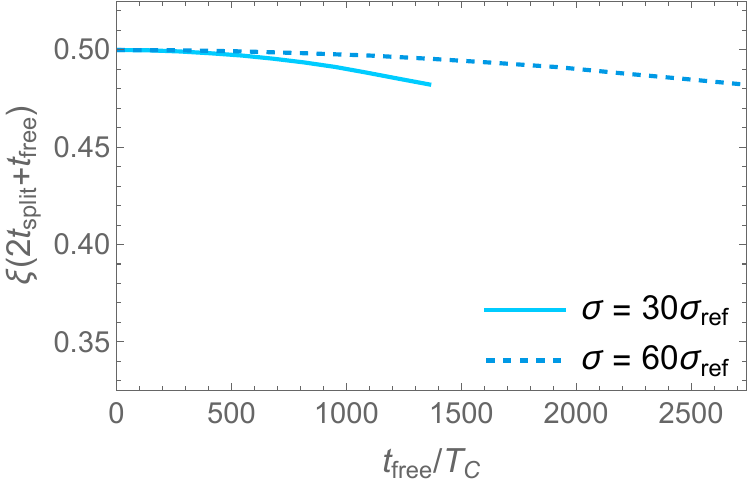}
    \caption{The graph exhibits~$\xi(2 t_\text{split} + t_\text{free})$ for a particle with mass~$m = M_\text{C} / 2$ ($\lambdabar = 2 L_\text{C}$) as a function of~$t_\text{free}$ for~$\sigma = 30 \; \sigma_\text{ref}$ (full line) and~$\sigma = 60 \; \sigma_\text{ref}$ (dashed line), assuming~$D = 100 \; \sigma^2 / L_\text{C}$ and~$L_\text{C} = L_\text{P}$ ($M_\text{C} = M_\text{P}$).}
    \label{fig_xi-time}
\end{figure}

\begin{figure}[b]
    \includegraphics[width = 80mm]{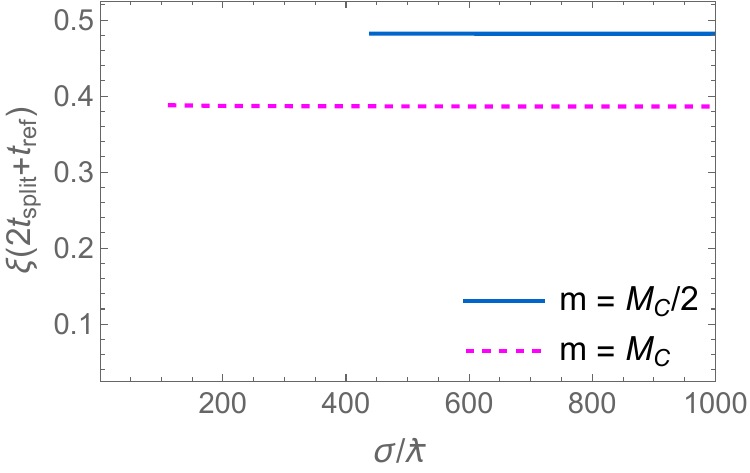}
    \caption{The graph exhibits~$\xi(2 t_\text{split} + t_\text{ref})$ as a function of~$\sigma / \lambdabar$ for~$m = M_\text{C} / 2$ ($\lambdabar = 2 L_\text{C}$, full line) and~$m = M_\text{C}$ ($\lambdabar = L_\text{C}$, dashed line), assuming~$D = 100 \; \sigma^2 / L_\text{C}$ and~$L_\text{C} = L_\text{P}$ ($M_\text{C} = M_\text{P}$). The larger the mass, the larger the loss of quantum correlations.}
    \label{fig_xi-sigma}
\end{figure}

\begin{figure}[t]
    \includegraphics[width = 80mm]{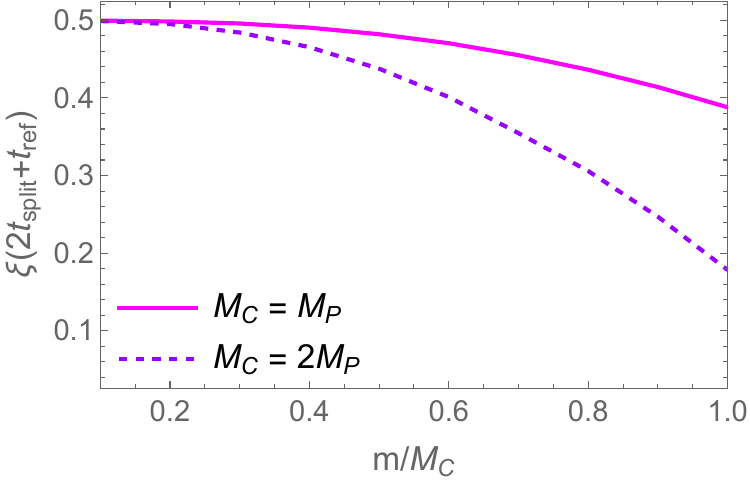}
    \caption{The graph exhibits~$\xi(2 t_\text{split} + t_\text{ref})$ as a function of~$m / M_\text{C}$ ($= L_\text{C} / \lambdabar$) assuming~$L_\text{C} = L_\text{P}$ ($M_\text{C} = M_\text{P}$, solid line) and~$L_\text{C} = L_\text{P} / 2$ ($M_\text{C} = 2 M_\text{P}$, dashed line). The larger the mass, the larger the loss of quantum correlations. (We have chosen~$\sigma = 30 \; \sigma_\text{ref}$ and~$D = 100 \; \sigma^2 / L_\text{C}$ for every value of~$m / M_\text{C}$.)}
    \label{fig_xi-mass}
\end{figure}

To quantify quantum correlations of the off-diagonal elements ($\zeta \neq \zeta'$), we define
\begin{eqnarray}
    \xi(2 t_\text{split} + t_\text{free})
    &\equiv&
    |\varrho_{\zeta \zeta'}(2 t_\text{split} + t_\text{free})|
    \nonumber \\
    &\equiv&
    |\langle \zeta | \Hat{\varrho}(2 t_\text{split} + t_\text{free}) | \zeta' \rangle|.
\end{eqnarray}
In Fig.~\ref{fig_xi-time} we plot~$\xi(2 t_\text{split} + t_\text{free})$ as a function of~$t_\text{free}$ for a particle with mass~$m = M_\text{C} / 2$ and two values of~$\sigma$. The curves are interrupted at~$t_\text{free} = t_\text{ref}$. The smaller the~$\sigma$, the faster the loss of quantum correlations (in comparison to the~QM~value of 0.5). (Error bars are negligibly small.) Figure~\ref{fig_xi-sigma} plots~$\xi(2 t_\text{split} + t_\text{ref})$ as a function of~$\sigma$ for distinct mass values. The larger the mass~$m$, the smaller~$\xi(2 t_\text{split} + t_\text{ref})$, and the smaller the quantum correlations. Next, Fig.~\ref{fig_xi-mass} plots~$\xi(2 t_\text{split} + t_\text{ref})$ as a function of~$m / M_\text{C}$ for different values of~$M_\text{C}$. The larger the mass, the larger the loss of quantum correlations. Figures~\ref{fig_xi-time}, \ref{fig_xi-sigma}, and~\ref{fig_xi-mass} are consistent with the decoherence exhibited in Figs.~\ref{fig_mu-time}, \ref{fig_mu-sigma}, and~\ref{fig_mu-mass}.

\begin{figure}[b]
    \includegraphics[width = 80mm]{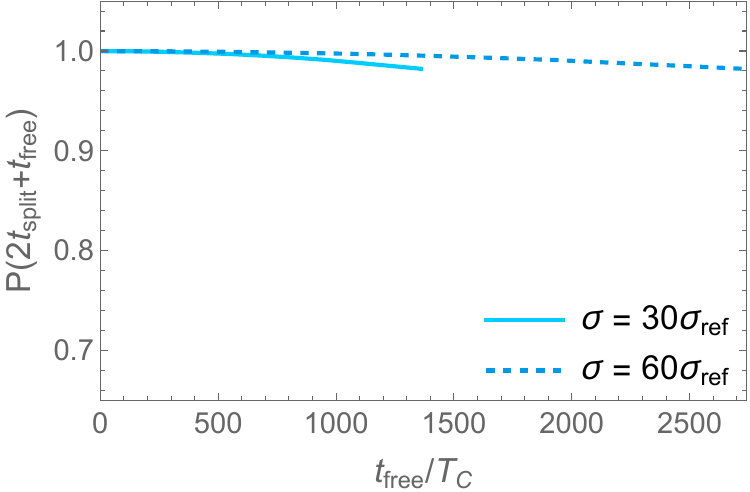}
    \caption{The probability~$P(2 t_\text{split} + t_\text{free})$ for a particle with mass~$m = M_\text{C} / 2$ ($\lambdabar = 2 L_\text{C}$) to have spin~$+ \hbar / 2$ along the $x$~axis is exhibited as a function of~$t_\text{free}$ for~$\sigma = 30 \; \sigma_\text{ref}$ (full line) and~$\sigma = 60 \; \sigma_\text{ref}$ (dashed line), assuming~$D = 100 \; \sigma^2 / L_\text{C}$ and~$L_\text{C} = L_\text{P}$ ($M_\text{C} = M_\text{P}$).}
    \label{fig_p-time}
\end{figure}

Finally, let us show how a spin measurement can distinguish our model from standard~QM~predictions. For this purpose, let us calculate the probability of obtaining~$+ \hbar / 2$ for a spin measurement along the $x$~axis:
\begin{equation}
    P(2 t_\text{split} + t_\text{free})
    \equiv
    \text{tr}[\Hat{\varrho}(2 t_\text{split} + t_\text{free}) | \rightarrow \rangle \langle \rightarrow |],
\end{equation}
where~$| \rightarrow \rangle$ is an eigenstate of the usual~$\Hat{S}_x$ operator: $\Hat{S}_x | \rightarrow \rangle \equiv (+ \hbar / 2) | \rightarrow \rangle$. In Fig.~\ref{fig_p-time} we plot~$P(2 t_\text{split} + t_\text{free})$ as a function of~$t_\text{free}$ for a particle with mass~$m = M_\text{C} / 2$ and two values of~$\sigma$. The curves are interrupted at~$t_\text{free} = t_\text{ref}$. The smaller the~$\sigma$, the faster~$P(2 t_\text{split} + t_\text{free})$ decreases with respect to the unity (predicted by~QM). Figure~\ref{fig_p-sigma} plots~$P(2 t_\text{split} + t_\text{ref})$ as a function of~$\sigma$ for distinct mass values. The larger the mass~$m$, the smaller~$P(2 t_\text{split} + t_\text{ref})$. Next, Fig.~\ref{fig_p-mass} plots~$P(2 t_\text{split} + t_\text{ref})$ as a function of~$m / M_\text{C}$ for different values of~$M_\text{C}$. As expected, the larger the mass, the more~$P(2 t_\text{split} + t_\text{ref})$ decreases.

{Figures~\ref{fig_p-time}, \ref{fig_p-sigma}, and~\ref{fig_p-mass} show how a double Stern-Gerlach experiment can distinguish our model from standard~QM~predictions. Interestingly, it would also distinguish our proposal from Schr{\" o}dinger-Newton equation models (since the latter and~QM give identical results for the initial state~\eqref{SV1}; see Refs.~\cite{2024-aguiar, 2024-grossardt}). Despite being a paramount challenge to keep massive quantum systems isolated from external influences, the fast progress in quantum technology supports our optimism~\cite{2021-gasbarri}.

\section{Conclusions}
\label{sec_conclusions}

\begin{figure}[t]
    \includegraphics[width = 80mm]{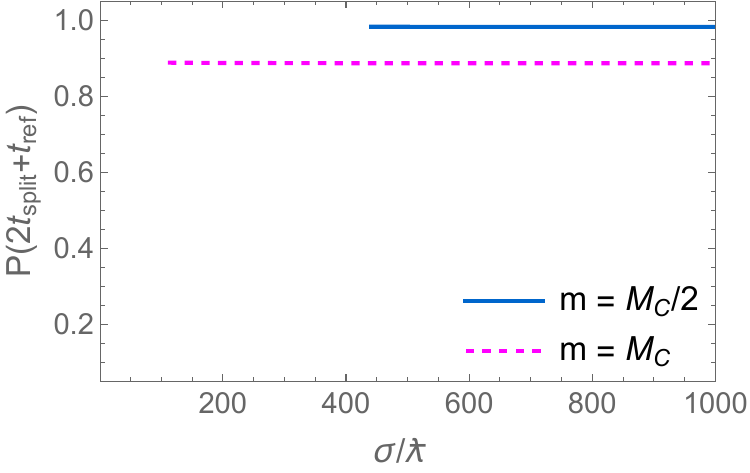}
    \caption{The probability~$P(2 t_\text{split} + t_\text{ref})$ for the spin to be~$+ \hbar / 2$ along the $x$~axis is exhibited as a function of~$\sigma / \lambdabar$ for~$m = M_\text{C} / 2$ ($\lambdabar = 2 L_\text{C}$, full line) and~$m = M_\text{C}$ ($\lambdabar = L_\text{C}$, dashed line), assuming~$D = 100 \; \sigma^2 / L_\text{C}$ and~$L_\text{C} = L_\text{P}$ ($M_\text{C} = M_\text{P}$). The larger the mass, the more the probability decreases.}
    \label{fig_p-sigma}
\end{figure}

\begin{figure}[t]
    \includegraphics[width = 80mm]{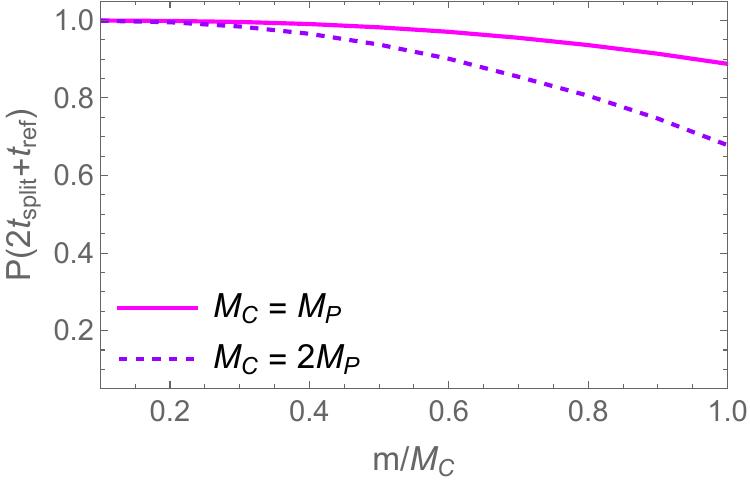}
    \caption{The probability~$P(2 t_\text{split} + t_\text{ref})$ for the spin to be~$+ \hbar / 2$ along the $x$~axis is exhibited as a function of~$m / M_\text{C}$ ($= L_\text{C} / \lambdabar$), assuming~$L_\text{C} = L_\text{P}$ ($M_\text{C} = M_\text{P}$, solid line) and~$L_\text{C} = L_\text{P} / 2$ ($M_\text{C} = 2 M_\text{P}$, dashed line). The larger the mass, the more the probability deviates from the standard~QM~prediction. (We have chosen~$\sigma = 30 \; \sigma_\text{ref}$ and~$D = 100 \; \sigma^2 / L_\text{C}$ for every value of~$m / M_\text{C}$.)}
    \label{fig_p-mass}
\end{figure}

We have raised the possibility that the {\em quantum-} and {\em classical-}particle concepts (anchored to classical spacetimes) independently emerge from some still unknown~QST, depending on whether the mass is {\em below} or {\em above} a certain Heisenberg-cut scale, respectively --- see Fig.~\ref{fig_manifesto}. In this framework, particles with masses~$m \sim M_\text{P}$ would be ruled by new physics. Although our effective model {\em cannot} be pushed to predict the behavior of macroscopic particles ($m / M_\text{C} \gg 1$), it is fair to expect that a full~QST would push forward the curves in Fig.~\ref{fig_eta-mass} to extremely small purity values for~$m / M_\text{C} \gg 1$, explaining the classical nature of the macroscopic world~\cite{1928-bohr}. The physical picture behind it is that coherence would leak from the particle to (non-observable) quantum degrees of freedom of the spacetime. This would be extremely inefficient for particles with~$m / M_\text{C} \ll 1$ but efficient for~$m / M_\text{C} \sim 1$. As discussed in Sec.~\ref{sec_sg}, a double Stern-Gerlach experiment followed by a spin measurement can distinguish our model from usual~QM. Also, the lack of free parameters in our model ($M_\text{C} \sim M_\text{P}$) leaves almost no room for amendments, if the model does not comply with the data (turning out beneficial for the sake of falsification). Although testing our model with current technology is extremely difficult~\cite{2019-fein, 2015-kovachy, 2021-margalit}, forthcoming technological improvements in matter-wave and Stern-Gerlach interferometry are much expected~\cite{2021-gasbarri, 2014-arndt, 2022-marshman}.

\acknowledgments

G.E.A.M. would like to acknowledge numerous discussions it has had with D. Vanzella and D. Sudarsky (among other colleagues) over the past many years. The authors also thank Juan P{\^e}gas for feedback. G.H.S.A. was fully supported by the S{\~ a}o Paulo Research Foundation (FAPESP) under grant~2022/08424-3. G.E.A.M. was partially supported by the National Council for Scientific and Technological Development and FAPESP under grants~301508/2022-4 and~2022/10561-9, respectively.

\appendix

\section{Derivation of Eq.~\eqref{MV1}}
\label{appendix}

In general, the mean square momentum of a particle is defined by
\begin{equation}
    \left.\langle \mathbf{p}^2 \rangle\right|_t
    \equiv
    \text{tr}[\hat{\rho}(t) \Hat{\mathbf{p}}^2]
    =
    \int \! d^3 \mathbf{p} \; \mathbf{p}^2 \langle \mathbf{p} | \Hat{\rho}(t) | \mathbf{p} \rangle,
    \label{MSM1}
\end{equation}
where~$\Hat{\rho}(t)$ is the particle's density matrix, and~$\Hat{\mathbf{p}}^2$ is the squared-momentum operator with~$\Hat{\mathbf{p}}^2 | \mathbf{p} \rangle \equiv \mathbf{p}^2 | \mathbf{p} \rangle$. For a particle-clone system described by a wavefunction~$\Psi(\mathbf{r}, \Bar{\mathbf{r}}, t)$, this matrix corresponds to
\begin{equation}
    \Hat{\rho}(t)
    =
    \int \! d^3 \mathbf{r} \int \! d^3 \mathbf{r}' \int \! d^3 \Bar{\mathbf{r}} \; \Psi(\mathbf{r}, \Bar{\mathbf{r}}, t) \Psi^*(\mathbf{r}', \Bar{\mathbf{r}}, t) | \mathbf{r} \rangle \langle \mathbf{r}' |,
\end{equation}
and Eq.~\eqref{MSM1} can be written as
\begin{eqnarray}
    \left.\langle \mathbf{p}^2 \rangle\right|_t
    &=&
    \int \! d^3 \mathbf{r} \int \! d^3 \mathbf{r}' \int \! d^3 \Bar{\mathbf{r}} \; \Psi(\mathbf{r}, \Bar{\mathbf{r}}, t) \Psi^*(\mathbf{r}', \Bar{\mathbf{r}}, t)
    \nonumber \\
    &{}&
    \times \frac{1}{(2 \pi \hbar)^3} \int \! d^3 \mathbf{p} \; \mathbf{p}^2 e^{i \mathbf{p} \cdot (\mathbf{r}' - \mathbf{r}) / \hbar}
    \nonumber \\
    &=&
    \int \! d^3 \mathbf{r} \int \! d^3 \Bar{\mathbf{r}} \; \Psi^*(\mathbf{r}, \Bar{\mathbf{r}}, t) [- \hbar^2 \nabla_\mathbf{r}^2 \Psi(\mathbf{r}, \Bar{\mathbf{r}}, t)],
    \nonumber \\
    &{}&
    \label{MSM2}
\end{eqnarray}
where we have used that~$\langle \mathbf{r}' | \mathbf{p} \rangle \equiv e^{i \mathbf{p} \cdot \mathbf{r}' / \hbar} / (2 \pi \hbar)^{3 / 2}$ in the first equality and
\begin{equation}
    \frac{1}{(2 \pi \hbar)^3} \int \! d^3 \mathbf{p} \; \mathbf{p}^2 e^{i \mathbf{p} \cdot (\mathbf{r}' - \mathbf{r}) / \hbar}
    =
    - \hbar^2 \nabla_\mathbf{r}^2 \delta^3(\mathbf{r}' - \mathbf{r})
\end{equation}
in the second one. For the particle-clone wavefunction~\eqref{S1}, it follows from Eq.~\eqref{MSM2} that
\begin{eqnarray}
    \left.\langle \mathbf{p}^2 \rangle\right|_t
    &=&
    \int \! d^3 \mathbf{r} \int \! d^3 \Bar{\mathbf{r}} \; \frac{\hbar^2}{(2 \pi \sigma^2)^3} \exp{\bigg(- \frac{\mathbf{r}^2 + \Bar{\mathbf{r}}^2}{2 \sigma^2}\bigg)}
    \nonumber \\
    &{}&
    \times \bigg\{\frac{3}{2 \sigma^2} + \frac{3 i a(\mathbf{r}, \Bar{\mathbf{r}}, t) L_\text{C}^2}{|\mathbf{r} - \Bar{\mathbf{r}}|_{L_\text{C}}^2} - \bigg[\frac{\mathbf{r}}{2 \sigma^2}
    \nonumber \\
    &{}&
    + i a(\mathbf{r}, \Bar{\mathbf{r}}, t) (\mathbf{r} - \Bar{\mathbf{r}})\bigg]^2\bigg\},
    \label{MSM3}
\end{eqnarray}
where
\begin{equation}
    a(\mathbf{r}, \Bar{\mathbf{r}}, t)
    \equiv
    \frac{\hbar t}{\lambdabar^2 |\mathbf{r} - \Bar{\mathbf{r}}|_{L_\text{C}}^3}.
\end{equation}
Here, it is convenient to define the coordinate transformations
\begin{equation}
    \begin{cases}
    x = \frac{X' + X}{2}
    \\
    \Bar{x} = \frac{X' - X}{2}
    \end{cases},
    \quad
    \begin{cases}
    y = \frac{Y' + Y}{2}
    \\
    \Bar{y} = \frac{Y' - Y}{2}
    \end{cases},
    \quad
    \begin{cases}
    z = \frac{Z' + Z}{2}
    \\
    \Bar{z} = \frac{Z' - Z}{2}
    \end{cases},
\end{equation}
\begin{equation}
    \begin{cases}
        X = R \sin{\theta} \cos{\phi}
        \\
        Y = R \sin{\theta} \sin{\phi}
        \\
        Z = R \cos{\theta},
    \end{cases},
    \quad
    \text{and}
    \quad
    \begin{cases}
        X' = R' \sin{\theta'} \cos{\phi'}
        \\
        Y' = R' \sin{\theta'} \sin{\phi'}
        \\
        Z' = R' \cos{\theta'}
    \end{cases},
\end{equation}
driving Eq.~\eqref{MSM3} to
\begin{widetext}
\begin{eqnarray}
    \left.\langle \mathbf{p}^2 \rangle\right|_t
    &=&
    \int\limits_0^\infty \! d R \int\limits_0^\pi \! d \theta \int\limits_0^{2 \pi} \! d \phi \int\limits_0^\infty \! d R' \int\limits_0^\pi \! d \theta' \int\limits_0^{2 \pi} \! d \phi' \; \bigg(\frac{1}{2}\bigg)^3 (R^2 \sin{\theta}) (R'^2 \sin{\theta'}) \frac{\hbar^2}{128 \pi^3 \sigma^{10}} e^{- (R^2 + R'^2)/(4 \sigma^2)} \bigg\{24 \sigma^2 - R^2 - R'^2
    \nonumber \\
    &{}&
    + \frac{2 i \alpha(R, t)}{R^2 + L_\text{C}^2} (6 \sigma^2 L_\text{C}^2 - R^2 L_\text{C}^2 - R^4) + R^2 \alpha^2(R, t) - 2 R R' [1 + i \alpha(R, t)] [\cos{\theta} \cos{\theta'} + \sin{\theta} \sin{\theta'} \cos{(\phi - \phi')}]\bigg\},
    \nonumber \\
    &{}&
    \label{MSM4}
\end{eqnarray}
\end{widetext}
where
\begin{equation}
    \alpha(R, t)
    \equiv
    \frac{4 \hbar \sigma^2 t}{\lambdabar^2 (R^2 + L_\text{C}^2)^{3 / 2}}.
\end{equation}
The integration in Eq.~\eqref{MSM4} can be directly calculated by Mathematica software, and the result is given in Eq.~\eqref{MV1}. (Note that~$\left.\langle \mathbf{p}^2 \rangle\right|_t = \left.\langle \Bar{\mathbf{p}}^2 \rangle\right|_t$ by construction.)

\end{document}